\documentclass{nature}
\usepackage{amsmath}
\usepackage[T1]{fontenc}
\usepackage[latin1]{inputenc}
\usepackage{graphicx}
\usepackage{amssymb}
\usepackage{dcolumn}
\usepackage{color}
\usepackage{bm}
\usepackage[normalem]{ulem}
\usepackage{url}
\usepackage{pdfpages}
\usepackage{pgffor}
\begin{document}

\title{Interaction driven giant thermopower in magic-angle twisted bilayer graphene}
\author{Arup Kumar Paul$^{1}$\footnote{equally contributed}, Ayan Ghosh$^{1}$\footnote{equally contributed}, Souvik Chakraborty$^{1}$\footnote{equally contributed}, Ujjal Roy$^{1}$, Ranit Dutta$^{1}$, K. Watanabe$^{2}$,T. Taniguchi$^{2}$, Animesh Panda$^1$, Adhip Agarwala$^{3}$, Subroto Mukerjee$^{1}$, Sumilan Banerjee$^{1}$ and Anindya Das$^{1}$\footnote{anindya@iisc.ac.in}}

\maketitle

\begin{affiliations}

\item Department of Physics,Indian Institute of Science, Bangalore, 560012, India.
\item National Institute for Materials Science, 1-1 Namiki, Tsukuba 305-0044, Japan.
\item Max Planck Institute for the Physics of Complex Systems,  N{\"o}thnitzer Stra\ss e 38, 01187 Dresden, Germany.

\end{affiliations}

\noindent\textbf{Magic-angle twisted bilayer graphene (MtBLG) has proven to be an extremely promising new platform to realize and study a host of emergent quantum phases arising from the strong correlations in its narrow bandwidth flat band. In this regard, thermal transport phenomena like thermopower, in addition to being coveted technologically, is also sensitive to the particle-hole (PH) asymmetry, making it a crucial tool to probe the underlying electronic structure of this material. We have carried out thermopower measurements of MtBLG as a function of carrier density, temperature and magnetic field, and report the observation of an unusually large thermopower reaching up to a value as high as $\sim \bf{100\mu V/K}$ at a low temperature of 1K. Surprisingly, our observed thermopower exhibiting peak-like features in close correspondence to the resistance peaks around the integer Moire fillings, including the Dirac Point, violates the Mott formula. 
We show that the large thermopower peaks and their 
associated behaviour arise from the emergent highly PH asymmetric electronic structure due to the cascade of Dirac revivals. Furthermore, the thermopower shows an anomalous peak around the superconducting transition on the hole side and points towards the possible role of enhanced superconducting fluctuations in MtBLG.}


\noindent\textbf{Introduction.}\
Interactions in many body 
systems lead to various complex emergent quantum phenomena like superconductivity, magnetism and correlated insulating phases. Understanding these many body quantum phenomena and utilizing their various applicability remains a key focus of condensed matter research. For this reason, MtBLG 
is a promising material with its flat band~\cite{Bistritzer12233,cao2018correlated,cao2018unconventional} induced plethora of exotic states like correlated insulator~\cite{cao2018correlated,yankowitz2019tuning,wu2021chern,saito2020independent,lu2019superconductors}, superconductivity~\cite{cao2018unconventional,yankowitz2019tuning,lu2019superconductors,arora2020superconductivity,saito2020independent,park_tunable_2021}, ferromagnetism~\cite{Sharpe605}, Chern insulator~\cite{wu2021chern,das_symmetry-broken_2021,nuckolls_strongly_2020,choi_correlation-driven_2021,stepanov2020competing,Andrew2021}, quantum anomalous Hall effect~\cite{Serlin900}, nematicity~\cite{jiang2019charge,cao_nematicity_2021} and Pomeranchuk effect~\cite{rozen2020entropic,saito2021isospin}. The discovery of these emergent quantum phases together with its easy tunability using a variety of experimental knobs makes MtBLG an unprecedented platform to probe the role of interactions in its unique electronic band-structure and further the search of novel electronic properties with technological applicability. In this direction, primarily electrical transport and local spectroscopic measurements have been utilised to probe and study the nature of the various symmetry-breaking electronic states~\cite{choi2019electronic,xie2019spectroscopic,kerelsky2019maximized,wong2020cascade,zondiner2020cascade}. Notably, recent measurements of local compressibility~\cite{zondiner2020cascade} and scanning tunneling microscopy~\cite{wong2020cascade} have revealed that the Fermi surface of MtBLG is highly malleable and undergoes interaction-driven quantum phase transitions at integer fillings of Moire lattice. The key finding is the resetting of the Fermi surface with strongly PH asymmetric density of states (DOS) around the integer fillings via a cascade of Dirac revival transitions~\cite{wong2020cascade,zondiner2020cascade}. However, their unambiguous signatures in global transport measurement are still lacking. Some signature 
is observed in Hall measurement~\cite{wu2021chern,park_tunable_2021}, where the Hall carrier density suddenly resets from finite value to zero without changing its sign at the integer fillings, but, the nature and degree of PH asymmetry of the electronic structure at the transition points remain unexplored.

\begin{figure*}
\centerline{\includegraphics[width=1\textwidth]{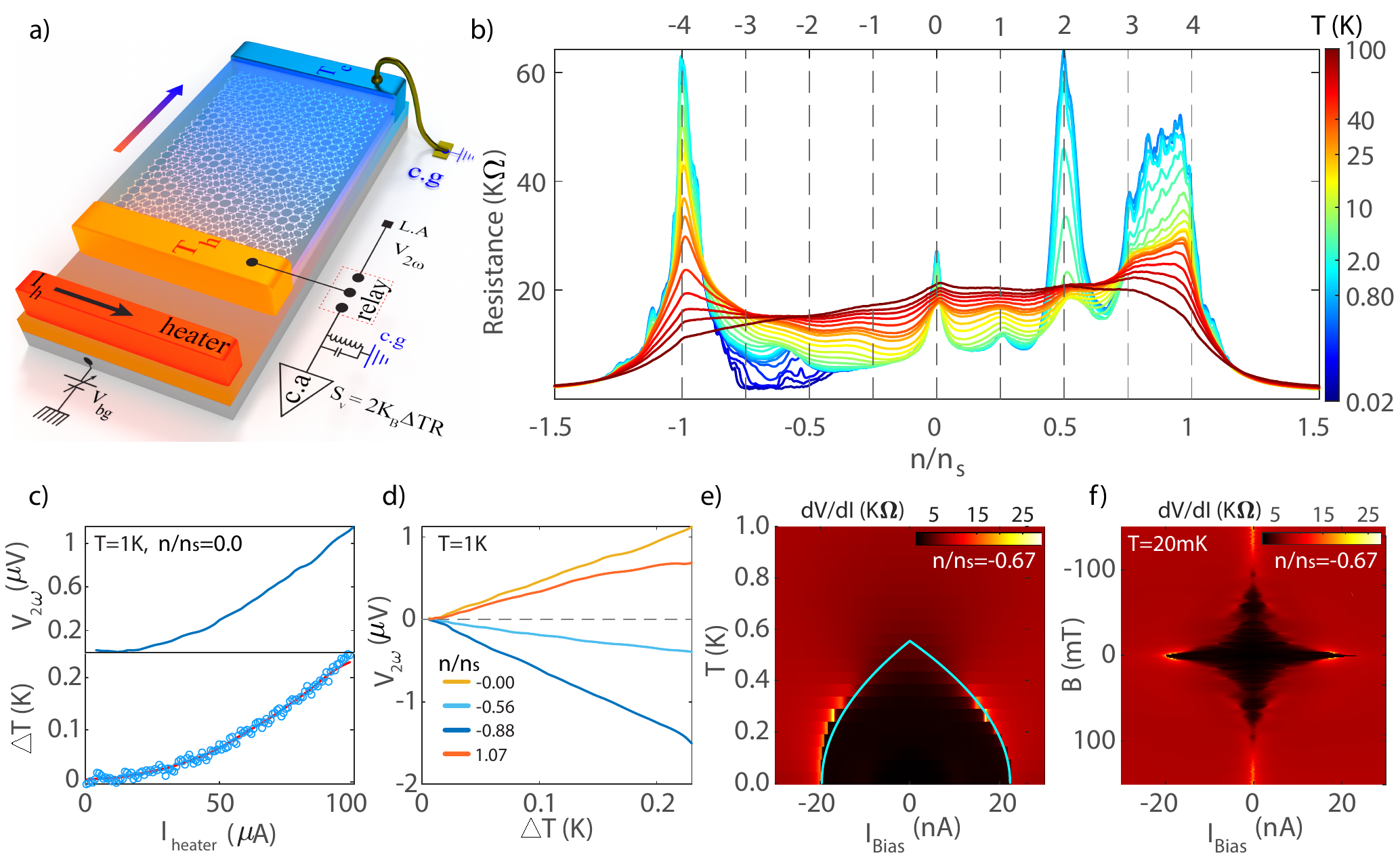}}
\caption{\textbf{Thermopower measurement set-up and device response.}  \textbf{a}, Set-up of devices. Passing a current, $I_h$ through the heater creates the temperature gradient across the device, where the colder end was directly bonded to the cold ground (c.g). The gate voltage, $V_{bg}$ controls the carrier density ($n$) of the device. The relay switches between the low-frequency and high-frequency measurement schemes. Low frequency, $2\omega$ method was used to measure the thermoelectric voltage ($V_{2\omega}$) at $\sim 13$ Hz using standard lock-in technique. High frequency ($\sim 720kHz$) thermal noise ($S_{V}$) measurement consisting LC resonant tank circuit and cryo-amplifier (CA) was used to measure the temperature difference ($\Delta T$) across the  device as $S_{V} = 2k_{B}\Delta T R$, where $R$ is the resistance of the device. \textbf{b}, Resistance versus filling fraction ($n/n_s$) as a function of increasing temperatures, where 
$n_{s}$ is the carrier density required to fill the flat band. The top axis shows in terms of numbers of electrons ($\nu$) per Moire lattice. The resistance peaks at the positive integer fillings ($\nu$) are visible at lower temperatures. At the hole side no such peaks are observed except at full filling, and bellow $\sim500mK$ resistance drops to $\sim 1.8k\Omega$, within $n/n_s -0.5$ to $-0.75$, which shows emergence of superconductivity. 
\textbf{c}, Measured $V_{2\omega}$ (upper panel) and $\Delta T$ (lower panel) as function of $I_h$ near the Dirac point at $1K$. \textbf{d} $V_{2\omega}$ as function of $\Delta T$ at $1K$, for different $n/n_s$, showing linear response regime. 
The slope of each curve gives the value of $S$. \textbf{e}, Differential resistance ($dV/dI$) versus bias current ($I_{sd}$) as function of temperature, at $n/n_s \sim -0.67$, where critical current ($I_c$) is maximum. The dark region corresponds to the superconducting region. The green solid line shows the theoretically generated $I_c$ using the BCS theory, $I_{c}(T) = I_{c}(0)(1-T/T_{c})^{2}$, where the $I_{c}(0)$ is the experimentally measured value at $T = 20mK$. 
\textbf{f}, $dV/dI$ with increasing perpendicular magnetic field at $n/n_s \sim -0.67$ and at $\sim 20mK$. The dark black region corresponds to the superconducting region, which is killed at $B_{\perp} = 0.1T$. 
}
\label{Figure1}
\end{figure*}

In this context, thermopower or the Seebeck effect is a unique 
tool to probe the PH asymmetry of the electronic structure of MtBLG. 
Compared to 
electrical transport, 
it is relatively non-invasive as an open circuit voltage ($\Delta V$) is measured across the sample in the presence of a small temperature gradient ($\Delta T$) relative to the sample temperature. In the linear regime, using semi-classical Boltzmann transport theory and assuming energy independent scattering time, the Seebeck coefficient ($S = -\Delta V/\Delta T$) can be written as $S=-(k_\mathrm{B}/Te)[\int (\epsilon-\mu)g(\varepsilon)(-df/d\varepsilon)d\varepsilon]/[\int g(\varepsilon)(-df/d\varepsilon)d\varepsilon]$, where $e$, $T$, $\mu$, $g(\varepsilon)$ and $-df/d\varepsilon$ are respectively the electronic charge, temperature, chemical potential, DOS and derivative of Fermi function. It can be seen that the numerator is an odd function due to the ($\epsilon-\mu$) term, and thus, the sign and magnitude of $S$ depend on nature and extent of asymmetry of the DOS around the chemical potential. 
As a result, $S$ is a highly sensitive probe to study the electronic structure around the 
transition points of MtBLG. 
Moreover, MtBLG with superconducting dome around half filling analogous 
of high-$T_c$ cuprate superconductors is an ideal playground to study the 
thermopower response 
as it has been employed to study the superconducting fluctuations in cuprates\cite{xu_vortex-like_2000}.

Motivated by these, we have extensively explored the thermopower response of MtBLG and non magic-angle tBLG devices. Unlike previous works involving graphene and tBLG~\cite{zuev2009thermoelectric,PhysRevB.80.081413,nam2010thermoelectric,wang2011enhanced,duan2016high,ghahari2016enhanced,mahapatra2020misorientation,ghawri2020excess}, we have utilized Johnson noise thermometry~\cite{Srivastaveaaw5798,fong2012ultrasensitive,crossno2016observation,betz2013supercollision} to directly measure the temperature gradient across the MtBLG device and accurately determine $S$ across a temperature ranging from $100mK$ to $10K$. Our measurements reveal intricate dependence of $S$ on carrier density ($\nu$), temperature ($T$), and magnetic field ($B$). Our key observations are following: i) The measured thermopower at low temperatures deviates completely from the expected zero-crossings following the semi-classical Mott formula~\cite{PhysRev.181.1336}. Instead, the thermopower exhibits peak-like features at all positive integer fillings including the Dirac point. ii) We observe a non-monotonic temperature dependence of the thermopower. The thermopower reaches a record high value of $\sim 100\mu V/K$ at $1K$ for half filling of the conduction band. iii) We also observe unusually large peaks in $S$ $\sim - (10-15)\mu V/K$ at sub-Kelvin temperatures around the superconducting transition tracing the superconducting dome in the hole side. We explain the first two results qualitatively using a simple model within self-consistent Hartree-Fock (HF) approximations showing emergent highly PH asymmetric DOS 
at integer fillings. Furthermore, we attribute the anomalous peaks around the $T_c$ to enhanced superconducting fluctuations due to inherent PH asymmetry in MtBLG. Our work highlights the ability of thermopower to independently provide unique insights into the novel quantum phenomena observed in MtBLG and opens a new route to achieve high thermoelectric cooling devices and generators at cryogenic temperatures.

\noindent\textbf{Set-up and device response.}\
Figure 1a shows the schematic of the device and the measurement setup for thermopower measurement. The devices consist of hBN encapsulated twisted bilayer graphene (tBLG) on a $Si/SiO_2$ substrate. 
The details are described in method and supplementary information (SI-1). 
For the thermopower measurement, an isolated gold heater line, as shown in Fig. 1a, is placed parallel to one side of the tBLG. 
To determine the thermopower or Seebeck coefficient ($S$), one needs to measure the generated thermoelectric voltage and the temperature difference ($\Delta T = T_h - T_c$). 
We have utilized well established $2\omega$ lock-in technique~\cite{zuev2009thermoelectric,PhysRevB.80.081413,nam2010thermoelectric,wang2011enhanced,duan2016high,ghahari2016enhanced,mahapatra2020misorientation} for measuring the thermoelectric voltage ($V_{2\omega}$) at $\omega \sim 13 Hz$. 
To measure $\Delta T$, we have utilized Johnson noise thermometry~\cite{Srivastaveaaw5798,fong2012ultrasensitive,crossno2016observation,betz2013supercollision}. 
The details of the noise thermometry setup can be found in our earlier works~\cite{Srivastaveaaw5798,PhysRevLett.126.216803} and shown in SI-3. The excess thermal noise, $S_V=2k_B\Delta TR$ measured across the sample is used to determine the $\Delta T$ (see SI-6 and SI-7), where $k_B$ is the Boltzmann constant and $R$ is the resistance of the device. Fig. 1c shows the measured $V_{2\omega}$ and the $\Delta T$ as a function of the heater current at a bath temperature ($T$) of 1K. In Fig. 1d, we plot $V_{2\omega}$ with $\Delta T$ for MtBLG 
at different carrier densities ($n$). The linearity of the plots in Fig. 1d suggests that we are in the linear regime, and the slope of each curve gives the $S$ for a given $n$. We have measured the $S$ from $100mK$ to $10K$ in the linear regime by adjusting the heater current such that the $\Delta T$ always remains much smaller than $T$ (SI-7). We have used three devices with twist angles of $\sim 0.26^0$, $1.05^0$ $1.86^0$.

The gate-dependent resistance ($R$) of MtBLG for different temperatures is shown in Fig. 1b. Here the gate voltage is replaced with an equivalent Moire filling factor, $\nu=4n/n_s$, where $n$ is the carrier density induced by the gate voltage and $n_s$ is the carrier density required to full-fill the flat band (4 electrons/holes per Moire unit cell). As can be seen from the $R$ versus $\nu$ response
, multiple resistance peaks appear at positive integer fillings, including the Dirac point, and these peak features survive up to $\sim$50K and above. On the contrary, for negative filling, we see the prominent resistance peak at $\nu=-4$, and between $\nu=-2$ and $\nu=-3$, 
the resistance drops 
below $600mK$ and saturates like a plateau 
at $\sim 1.8 k\Omega$ 
showing the emergence of superconductivity. 
The resistance value at the full-filling ($\nu=\pm 4$) continuously decreases with increasing temperature. 
On the other hand, resistance value at $\nu=0$ and $2$ decreases with increasing temperature up to $\sim 10 K$ and then increases linearly, showing metallic nature (SI-10 and SI-11). 
Fig. 1e and 1f plot the evolution of differential resistance ($\frac{dV}{dI}$) versus bias current ($I_{sd}$) response with temperature and perpendicular magnetic field, respectively, at $\nu \sim -2.5$, and confirms the existence of the superconductivity though the resistance is measured in two-probe geometry (Method and SI-12). 

\begin{figure*}
\centerline{\includegraphics[width=1\textwidth]{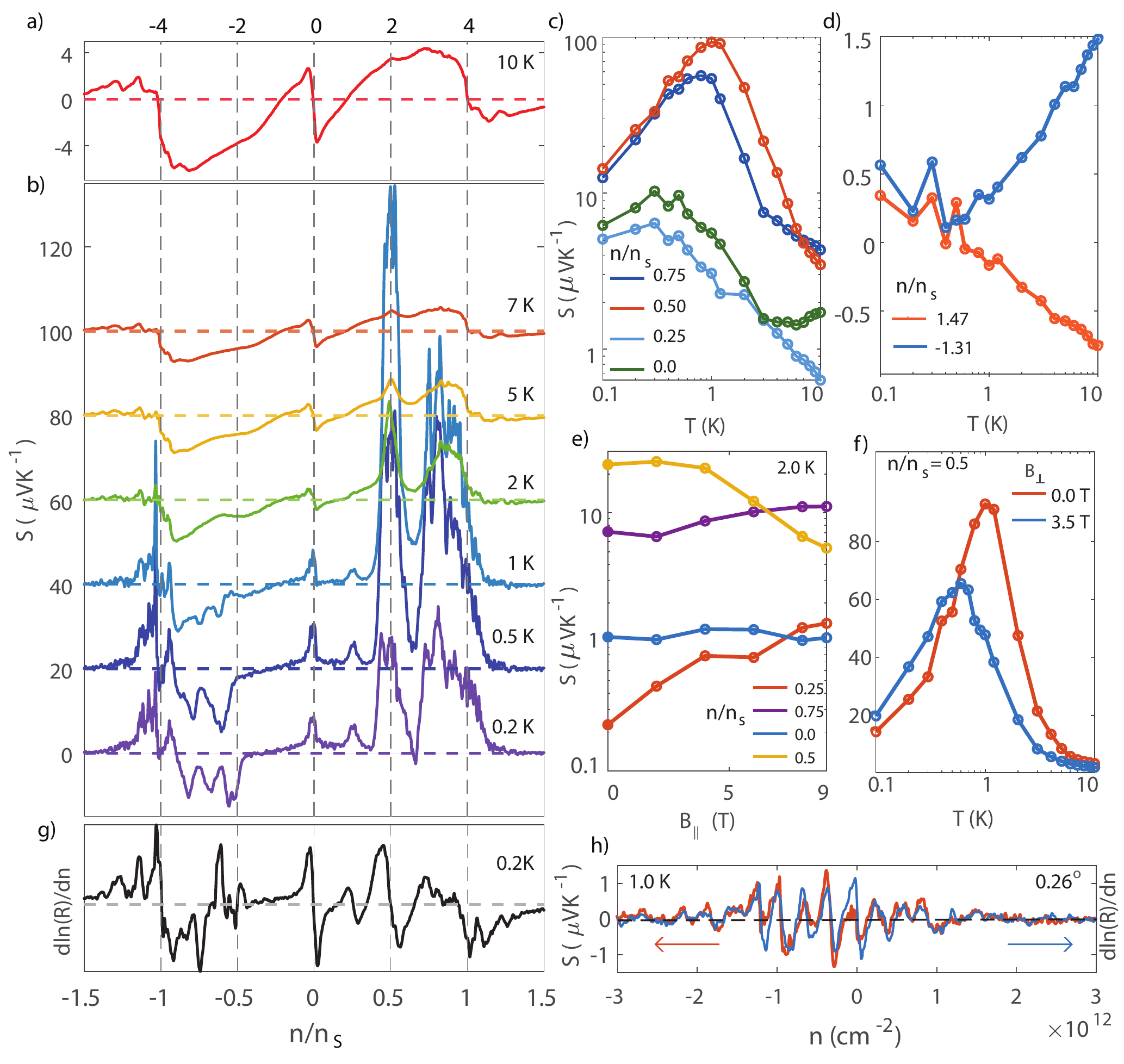}}
\caption{\textbf{Thermopower response at integer Moire fillings.} \textbf{a}, Measured thermopower of MtBLG with carrier fillings at $10K$. The dashed horizontal line corresponds to zero thermopower and vertical dashed lines correspond to the integer number of electron fillings of the Moire lattice. \textbf{b}, Measured thermopower at several temperatures from $0.2K$ to $7K$. For the clarity the data are shifted by $20\mu V/K$ along the y-axis. The dashed horizontal lines are the zero thermopower for the corresponding temperatures. \textbf{c}, Temperature dependence of $S$ for the flat band in log-log scale at $n/n_{s} = 0, 0.25, 0.5, 0.75$, and \textbf{d}, for dispersive bands at hole-side ($n/n_{s} = -1.31$) and electron-side ($n/n_{s} = 1.47$). \textbf{e}, $S$ with $B_{\parallel}$ for a different thermal cycle at $2K$. \textbf{f}, Comparison of $S$ at half-filling with temperature between zero and $B_{\perp}$ = $3.5T$. \textbf{g}, Derivative of resistance of MtBLG with carrier density at $0.2K$ according to Mott formula. The dashed horizontal line correspond to zero derivative line. The clear sign changes at $\nu = 0, 1, 2$ are seen but absent in the $S$ data in Fig. 2b. \textbf{h}, Measured $S$ and derivative of resistance for $0.26^{0}$ non-magic angle tBLG at $1K$. The qualitative agreement of sign changes between them can be seen.}
\label{Figure2}
\end{figure*} 

\noindent\textbf{Band reconstruction of MtBLG probed by thermopower.}\ 
The Fig. 2a and 2b show the measured thermopower versus $\nu$ at several temperatures, from $200mK$ to $10 K$ for MtBLG. 
At $10K$ (Fig. 2a), thermopower has approximate mirror symmetry for both conduction and valence band albeit with opposite signs. It can be seen that the thermopower changes its sign at the Dirac point, at flat band full-filling ($\nu \sim \pm 4$) and around $\nu \sim \pm 1$. The sign of the thermopower depends on the type of the carriers, positive for hole-like and negative for electron-like carriers, and its magnitude goes to zero at the symmetric points of the electronic structure as described by the semi-classical equation. Like, at the Dirac point, the density of states (DOS) goes to zero symmetrically from both the conduction and valence band. Similarly, at the band full-filling with the energy gap between the flat and higher energy-dispersive bands, the $S$ is expected to change the sign. One more sign change is expected at the middle of the conduction or valence band as the single-particle DOS of the flat band 
reaches a maximum (van-hove singularity - VHS) around $\nu=\pm 2$. 
If the DOS is symmetric around the maxima, one would expect a sign change in $S$ exactly at $\nu=\pm 2$. However, the inherent asymmetry of the DOS, which is complex for MtBLG, of the conduction band or valence band can give rise to the sign change shifted from $\nu=\pm 2$. 

As we decrease the temperature below $10K$, the apparent asymmetry of the $S$ (Fig. 2b) between the conduction and valence band grows similar to the asymmetry observed in the resistance data in Fig. 1b. Most importantly, the thermopower exhibits a positive peak around $\nu \sim 2$, and its magnitude increases rapidly with decreasing temperature and reaches a maximum value of $\sim 95-100\mu V/K$ at $\sim 1K$, followed by a decrement of the magnitude with a further reduction of the temperature. Similar, positive peaks are also seen around $\nu \sim 1,3$ at $\sim 2K$ and at the Dirac point below $1K$. 
The observed positive peak in thermopower at the positive integer fillings, including the Dirac point, is quite striking. Any energy gap $\gtrsim k_\mathrm{B}T$ either from the single-particle band structure or induced by electronic interactions will give a sign change of $S$. In particular, one would expect $S$ to go to zero at the resistance maxima, i.e., at the integer fillings as the Mott formula~\cite{PhysRev.181.1336} $S=(\pi^2 k_B T/3e)(d\ln(R)/dn)g(\epsilon)$ 
gives zero at those points and shown in Fig. 2g for $0.2K$. 
Thus, one can see a complete violation of the Mott formula for MtBLG. The violation persists even up to $10K$, as shown in SI-13. On the contrary, for non magic-angle tBLG devices, the measured sign of $S$ and the Mott formula matches well, as shown in Fig. 2h for $\sim 0.26^0$ (SI-Fig. 15b for $1.86^0$).

The recurring thermopower peaks (Fig. 2b) at integer fillings with a positive sign (which usually occurs for hole-like carriers) suggest, at least within an effective single-particle picture, repeated restructuring of the Fermi surface at integer fillings such that overall hole-like carriers are dominant. The pliable Fermi surfaces due to interactions around the integer fillings have been reported in MtBLG and Stoner like transitions~\cite{wong2020cascade,zondiner2020cascade,wu2021chern,park_tunable_2021} have been observed experimentally. The key features of these transitions are -- a Lifshitz transition followed by a Dirac revival, which essentially gives rise to large asymmetric DOS around the transition point such that, for $\nu>0$, from one side (left side of the transition), the DOS rapidly drops whereas other side (right side of the transition) the DOS increases gradually, similar to a sawtooth. Such asymmetric DOS can give rise to peak in $S$ with large value around the transition point as discussed in the theoretical section. It can be seen in Fig. 2h and SI-13 that for non magic-angle 
tBLG devices, we do not observe any thermopower peaks, and the measured $S$ is around $\sim 1\mu V/K$ at $\sim 1K$ as expected for graphene-based devices at such low temperatures~\cite{zuev2009thermoelectric,PhysRevB.80.081413,nam2010thermoelectric}.

The temperature dependence of $S$ for different integer fillings, including Dirac point, is shown in Fig. 2c. The common key feature is the non-monotonic temperature dependence of $S$ with a maxima at a certain temperature, which depends on the fillings. For example, at $\nu \sim 2$ and $\sim 3$ the peak appears around $\sim 1K$ whereas it is $\sim 0.3K$ for $\nu \sim 1$ and the Dirac point. The deviation from the linear $T$ dependence of $S$ again suggests the strong violation of Mott's formula~\cite{PhysRev.181.1336} for the flat band of MtBLG. However, it can be seen that for the dispersive bands ($n/n_{s} \sim 1.47$ and $n/n_{s} \sim -1.31$) the $S$ increases almost linearly with $T$ (Fig. 2d), consistent with the Mott formula. 
Furthermore, the response of $S$ with in-plane magnetic field ($B_{\parallel}$) underlies the nature of the ground states at different integer fillings. As can be seen in Fig. 2e (for a different thermal cycle as shown in SI-14) the thermopower peaks increase with $B_{\parallel}$ at $\nu \sim 1$ 
, but decreases at $\nu \sim 2$. 
These observations are consistent with the cascade of Dirac revival picture in Ref~\cite{zondiner2020cascade}, where the emergence of flavored symmetry breaking in MtBLG with 
polarized ground state at $\nu \sim 1$ 
strengthens the transition, and thus make it more PH asymmetric DOS resulting in higher $S$. At $\nu \sim 2$ the value of $S$ decreases with both $B_{\parallel}$ and $B_{\perp}$, as shown as a function of $T$ in Fig. 2f. It can be noticed that the peak position of the $S$ shifted to lower temperature $\sim 0.6K$ at $B_{\perp} = 3.5T$ with a value of $\sim 70 \mu V/K$, thus emphasizing its tunability for use as thermoelectric cooling devices at cryogenic temperatures. It should be noted that the $S$ of the dispersive band and non magic-angle tBLG devices remain insensitive to $B_{\parallel}$ (see SI-14).

\begin{figure*}
\centerline{\includegraphics[width=1\textwidth]{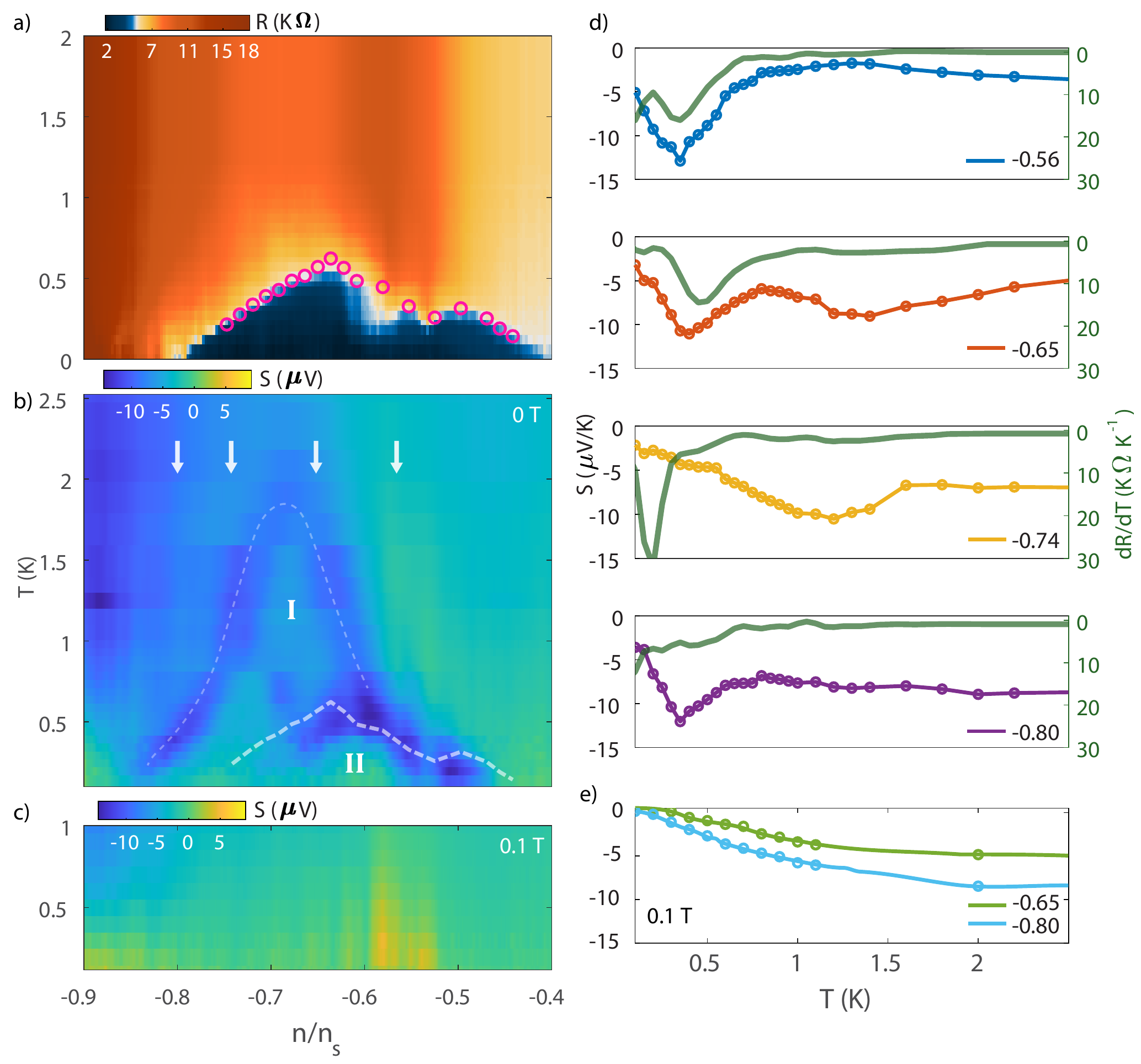}}
\caption{\textbf{Thermopower across the superconducting transition.} \textbf{a}, 2D colourmap of resistance as a function of temperature and carrier filling for the hole side flat band of MtBLG. The dark blue portion corresponds to the superconducting region around the weaker Mott peak at $n/n{s} \sim -0.55$. The open circles are the $T_{c}$ at different carrier densities as determined from the differential resistance versus critical current plot as a function of temperature as shown in Fig. 1e. \textbf{b}, 2D colourmap of measured thermopower as a function of temperature and carrier filling. The blue dark portions are the negative thermopower peak as shown as the cut lines in Fig. 3d for the carrier densities marked by the white vertical arrows. The regions $I$ and $II$ correspond to two different dome like portions, where region $II$ matches well with the superconducting dome seen in Fig. 3a as shown by the white dashed line, which is the trace of $T_{c}$ as shown in Fig.  3a. The other white dashed line enclosing the regions $I$ is the guiding line. \textbf{c}, The measured thermopower at a $B_{\perp} = 0.1T$ without peak like features in $S$. \textbf{d}, Open circles with the connected lines in different panels are the cut lines of measured $S$ as a function of temperature from Fig. 3b at $n/n_{s} = -0.56, -0.65, -0.74$ and $-0.8$. The solid lines are corresponding cut lines of $dR/dT$ from Fig. 3a. \textbf{e}, Thermopower cut lines for $B_{\perp} = 0.1T$ at $n/n_{s} = -0.65.$ and $-0.8$.}
\label{Figure3}
\end{figure*} 

\noindent\textbf{Anomalous thermopower response around the superconducting dome.}\ 
As shown in Fig. 2b, there are no thermopower peaks for the valence flat band of the MtBLG device in the temperature range of $2-10K$. This is consistent with the weaker cascaded transitions observed for the hole side in Ref~\cite{zondiner2020cascade}. It was shown that a larger twist angle $\sim 1.1^0$ is required to observe the sawtooth behavior in the DOS for the hole side. However, for the MtBLG device, some peak-like features are developed within $\nu \sim -2$ to $-3.5$ below $2K$. The number of peaks and their positions in filling changes with decreasing temperature; one weak negative peak is seen at $2K$, whereas three prominent negative peaks are observed at $0.2K$. Fig. 3a shows the two-dimensional (2D) color map of the resistance with temperature and fillings. The darker blue region corresponds to the superconducting phase seen close to a weak resistance peak at $n/n_{s} \sim -0.55$ (details in SI-12). The open circles denote superconducting transition temperature $T_{c}$ as a function of filling, obtained from critical current measurement shown in Fig. 1e and SI-12.

Fig. 3b shows the corresponding 2D color map of $S$ at zero magnetic fields. Here the darker blue shaded ribbons enclosing dome-like structures correspond to negative thermopower peaks. A closer view shows two inter-penetrating dome-like structures marked by the dashed white lines and labeled as $I$ and $II$ in Fig.3b. The dashed white line enclosing the region $II$ is the trace of $T_{c}$ as shown in Fig.3a, whereas the white dashed line enclosing the region $I$ is a guide to the eye to follow the locus of the broad negative thermopower peaks. As evident from Fig. 3a and 3b, there are clear resemblances between the peak position in $S$ enclosing region $II$ and $T_{c}(n)$ dome. This can be further ascertained from Fig. 3d, where the $S$ from Fig. 3b and $dR/dT$ for Fig. 3a are plotted as a function of $T$ for fixed fillings, $n/n_{s} = -0.56$ and $-0.65$. Both $S$ and $dR/dT$ exhibits prominent peaks around $T_c$. However, peak in $S$ enclosing region $I$ in Fig. 3b has hardly any direct correspondence in the resistance data in Fig. 3a, as can be seen in Fig. 3d, where $S(T)$ exhibits broad peak at $n/n_{s} = -0.74$ and $-0.80$, but $dR/dT$ does not show any such feature around the same temperature. 

Apart from the correlation between locus of the thermopower peak in the $n-T$ plane and $T_c(n)$ over a large part of the superconducting dome, the most important clue for the possible origin of the unusual thermopower peak is obtained by applying a $B_{\perp}$. As shown in Fig. 3c, the peak in $S$ completely disappears with the application of tiny $B_\perp=0.1$T. This is demonstrated in Fig. 3e by plotting $S(T)$ for $n/n_{s} = -0.65$ and $-0.8$. 
These observations suggest that anomalous peak in $S$ in Fig. 3b, particularly enclosing the region $II$, directly relates to the superconductivity of MtBLG. One cannot help but to notice the superficial resemblance of the trace of the broad thermopower peak enclosing region $I$ with the putative pseudogap temperature line in the doping-temperature plane of cuprates~\cite{timusk_pseudogap_1999}. However, the origin of the boundary enclosing region $I$ remains unclear at this moment and will be an interesting direction for future studies, like the possible role of nematicity and competing orders in MtBLG~\cite{cao_nematicity_2021}. Nonetheless, even the observed anomalous thermopower peak around $T_{c}$ is quite striking. Anomalously large thermopower response around $T_{c}$, and even extending far above $T_{c}$, have been reported in transverse thermoelectric coefficient, i.e. the Nernst coefficient~\cite{xu_vortex-like_2000}, for high $T_{c}$ cuprate superconductors. However, experimental observations of peaks in the longitudinal component of $S$ are scarce \cite{howson_anomalous_1989} and remain controversial. Such thermopower peaks have been theoretically predicted~\cite{howson_anomalous_1989,lu_fluctuation_1995} to exist above, albeit close to, $T_c$ from superconducting fluctuations under certain situations, like for a superconductor in the dirty limit~\cite{howson_anomalous_1989,lu_fluctuation_1995}. 
The observed thermopower in our MtBLG device might have similar origin, as disorder due to twist-angle inhomogeneity~\cite{zondiner2020cascade,uri_mapping_2020} is naturally present in the system. Effects of superconducting fluctuations in the thermopower~\cite{howson_anomalous_1989,lu_fluctuation_1995} are expected to be much more enhanced in MtBLG near half filling due to inherent PH asymmetry from the low-energy VHS of the flat band as well as the emergent PH asymmetry due to the proximate weak cascade transition in the hole side.


\begin{figure*}
\centerline{\includegraphics[width=1.0\textwidth]{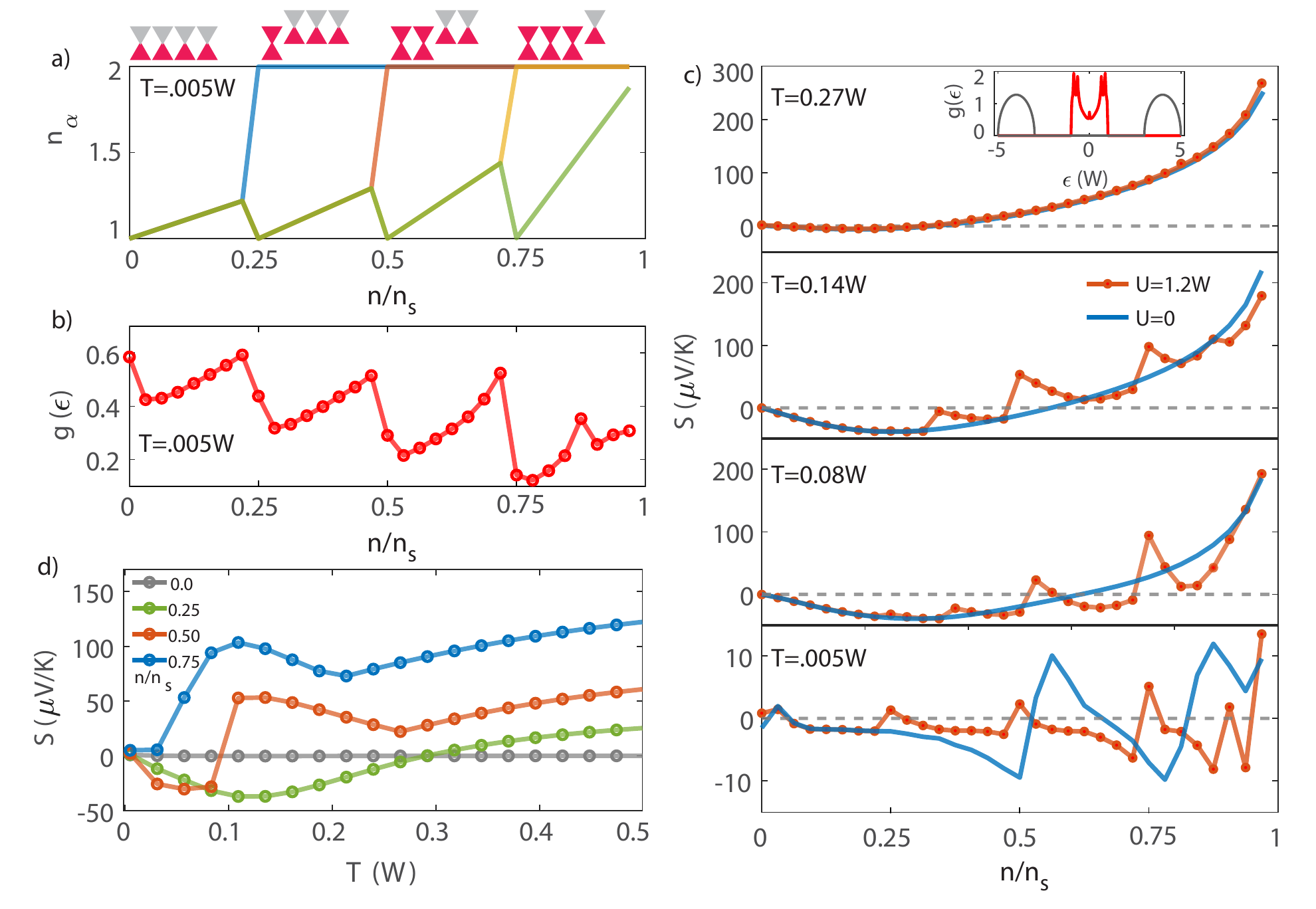}}
\caption{\textbf{Cascade of Dirac revivals and thermopower peaks around integer fillings.} \textbf{a}, The occupation ($n_\alpha,~\alpha=1,\dots,4$) of individual flavors as a function  filling $n/n_s$, obtained from Hartree-Fock (HF) calculations for $T=0.005W$ and a local inter-flavor interaction $U=1.2W$. At zero filling, the Dirac cones corresponding to the four spin-valley degrees of freedom are degenerate. A cascade of Stoner-like transitions close to the integer fillings lead to complete filling of one, two and three of the flavors successively while the filling of the remaining flavor(s) resets to Dirac point. \textbf{b}, The resultant HF DOS $g(\epsilon)$ at the chemical potential at $T=0.005W$ exhibits the sawtooth feature. The effective single-particle DOS $g(\epsilon)$ changes drastically at each integer fillings and shows strong low-energy particle-hole asymmetry (SI-Fig. 20). \textbf{c}, The calculated thermopower $S$ in HF approximation (red circles with line) shows peak-like features around the integer fillings due to the Dirac revivals (Fig. {\bf a}). The top to bottom panels are in the order of decreasing temperatures $T=0.27W,0.14W,0.08W,0.005W$. The thermopower $S_0$ obtained using 'rigid' non-interacting single particle DOS (Fig. {\bf c}, top panel, inset) is shown by the solid blue lines. The non-interacting $S_0$ exhibits one sign change around the half filling for higher temperature $T\gtrsim 0.08W$, and multiple sign changes across the two VHSs 
of the non-interacting DOS at very low temperature (the bottom panel), but these peaks depend on the details of non-interacting DOS and are not necessarily tied to integer the fillings, unlike $S(\nu)$ in the interacting cases. \textbf{d}, Non-monotonic temperature dependence of thermopower at integer fillings from the cascaded transitions. The Dirac revived symmetry broken state at $n/n_s=0.50$ only gets stabilized at finite temperature for the particular non-interacting DOS, as indicated by a sign change in $S$ around $T\sim 0.1W$.}
\label{Figure4}
\end{figure*} 

\noindent\textbf{Emergent low-energy particle-hole asymmetry and giant thermopower peaks.}\ 
As already mentioned, the thermopower peaks suggest strong emergent low-energy 
PH asymmetry of the putative correlated states at integer fillings, at least, within the effective single-particle or Hartree-Fock descriptions of various possible symmetry broken states \cite{PhysRevX.8.031089,PhysRevX.10.031034,shavit2021theory}. As discussed in method and SI-16, we use a simple minimal model \cite{zondiner2020cascade} with four fermionic flavors, corresponding to the spin and valley degrees of freedom, each described in terms of a single-particle DOS \cite{zondiner2020cascade,Bistritzer12233,PhysRevX.8.031087}, and interacting via a local Coulomb interaction. We treat the latter via self-consistent Hartree-Fock (HF) approximation and use the resulting HF DOS to calculate the resistivity and thermopower as a function of filling and temperature, using Kubo formulae (SI-16). We have used different non-interacting DOSs, obtained from both effective continuum models \cite{Bistritzer12233,PhysRevX.8.031087}, with and without lattice relaxation effects \cite{PhysRevX.8.031087}, as well as tight-binding model \cite{PhysRevB.85.195458} (SI-16). The main results are summarized in Fig. 4, where the peak value of $S$ reaches $\sim 50-100~\mu\mathrm{V/K}$ for $\nu\simeq 2,3$ at $T\sim 0.1 W$, consistent with our experimental observations (Fig. 2b). The temperature range $T\simeq 0.005W-0.27W$ corresponds to $\sim 200~\mathrm{mK}-13~\mathrm{K}$, for a bandwidth $2W\simeq 10~\mathrm{meV}$. For comparison, in Fig. 4c (solid blue lines), we have shown the $S_0$ for the non-interacting case (see Fig. 4c caption for details). We find the thermopower peak around an integer filling to be a robust feature whenever the Dirac revival is stabilized within the HF approximation, and support the simultaneous presence of thermopower (Fig. 2b) and resistance (Fig. 1b) peaks, as well as the non-monotonic temperature dependence of $S$ (Fig. 2c) in our experiment.


\noindent\textbf{Discussion.}\ 
Our theory qualitatively captures the thermopower peaks, but the $S(n)$ follows an overall `background' profile dictated by the non-interacting $S_0(n)$ (Fig. 4c) and its sign change around the half filling. There could be several reasons behind the deviation of $S$ obtained from HF approximation compared to the experimental one, e.g., effects of more complex and realistic single-particle DOS for MtBLG than the used continuum model~\cite{zondiner2020cascade,Bistritzer12233,PhysRevX.8.031087}, twist angle inhomogeneity~\cite{zondiner2020cascade,uri_mapping_2020} and strong correlations in the strange metal state~\cite{PhysRevLett.124.076801} (see SI-16 for a detailed discussion). Moreover, we should note that there are theoretical models~\cite{PhysRevX.8.031089,PhysRevX.10.031034,shavit2021theory} which lead to a small gap ($\Delta$) at the Dirac revivals. This will be consistent with the simultaneous presence of thermopower, and resistance peaks at integer fillings provided $k_\mathrm{B}T\gtrsim \Delta$. At very lower temperatures, $S$ is expected to change sign across the position of resistance peak. Thus, our thermopower results put a tighter upper bound, $\Delta \sim 0.1-0.2$ meV (activation gap in SI-11), on the correlation-induced gap at integer fillings. In the SI-16, we also discuss the expected thermopower from various other kinds of ground states 
and their possible signatures in our measurements.


\noindent\textbf{Conclusion.}\ 
In summary, our experiments reveal unusual low-temperature thermopower response and anomalously large peaks in Seebeck coefficient originating from emergent highly particle-hole asymmetry in the flat band of MtBLG. 
In the hole side of the MtBLG flat band, close to half-filling, our measurements also reveal two interpenetrating dome-like structures, traced by peaks in the thermopower on the doping-temperature plane. The boundary of the dome at lower temperature coincides with the line of superconducting transitions, presumably providing a rare glimpse of the enhanced superconducting fluctuations in MtBLG. 

\newpage

\bibliography{ref}

\begin{thebibliography}{10}
\expandafter\ifx\csname url\endcsname\relax
  \def\url#1{\texttt{#1}}\fi
\expandafter\ifx\csname urlprefix\endcsname\relax\def\urlprefix{URL }\fi
\providecommand{\bibinfo}[2]{#2}
\providecommand{\eprint}[2][]{\url{#2}}

\bibitem{Bistritzer12233}
\bibinfo{author}{Bistritzer, R.} \& \bibinfo{author}{MacDonald, A.~H.}
\newblock \bibinfo{title}{Moir{\'e} bands in twisted double-layer graphene}.
\newblock \emph{\bibinfo{journal}{Proceedings of the National Academy of
  Sciences}} \textbf{\bibinfo{volume}{108}}, \bibinfo{pages}{12233--12237}
  (\bibinfo{year}{2011}).

\bibitem{cao2018correlated}
\bibinfo{author}{Cao, Y.} \emph{et~al.}
\newblock \bibinfo{title}{Correlated insulator behaviour at half-filling in
  magic-angle graphene superlattices}.
\newblock \emph{\bibinfo{journal}{Nature}} \textbf{\bibinfo{volume}{556}},
  \bibinfo{pages}{80--84} (\bibinfo{year}{2018}).

\bibitem{cao2018unconventional}
\bibinfo{author}{Cao, Y.} \emph{et~al.}
\newblock \bibinfo{title}{Unconventional superconductivity in magic-angle
  graphene superlattices}.
\newblock \emph{\bibinfo{journal}{Nature}} \textbf{\bibinfo{volume}{556}},
  \bibinfo{pages}{43--50} (\bibinfo{year}{2018}).

\bibitem{yankowitz2019tuning}
\bibinfo{author}{Yankowitz, M.} \emph{et~al.}
\newblock \bibinfo{title}{Tuning superconductivity in twisted bilayer
  graphene}.
\newblock \emph{\bibinfo{journal}{Science}} \textbf{\bibinfo{volume}{363}},
  \bibinfo{pages}{1059--1064} (\bibinfo{year}{2019}).

\bibitem{wu2021chern}
\bibinfo{author}{Wu, S.}, \bibinfo{author}{Zhang, Z.},
  \bibinfo{author}{Watanabe, K.}, \bibinfo{author}{Taniguchi, T.} \&
  \bibinfo{author}{Andrei, E.~Y.}
\newblock \bibinfo{title}{Chern insulators, van {Hove} singularities and
  topological flat bands in magic-angle twisted bilayer graphene}.
\newblock \emph{\bibinfo{journal}{Nature Materials}}
  \textbf{\bibinfo{volume}{20}}, \bibinfo{pages}{488--494}
  (\bibinfo{year}{2021}).

\bibitem{saito2020independent}
\bibinfo{author}{Saito, Y.}, \bibinfo{author}{Ge, J.},
  \bibinfo{author}{Watanabe, K.}, \bibinfo{author}{Taniguchi, T.} \&
  \bibinfo{author}{Young, A.~F.}
\newblock \bibinfo{title}{Independent superconductors and correlated insulators
  in twisted bilayer graphene}.
\newblock \emph{\bibinfo{journal}{Nature Physics}}
  \textbf{\bibinfo{volume}{16}}, \bibinfo{pages}{926--930}
  (\bibinfo{year}{2020}).

\bibitem{lu2019superconductors}
\bibinfo{author}{Lu, X.} \emph{et~al.}
\newblock \bibinfo{title}{Superconductors, orbital magnets and correlated
  states in magic-angle bilayer graphene}.
\newblock \emph{\bibinfo{journal}{Nature}} \textbf{\bibinfo{volume}{574}},
  \bibinfo{pages}{653--657} (\bibinfo{year}{2019}).

\bibitem{arora2020superconductivity}
\bibinfo{author}{Arora, H.~S.} \emph{et~al.}
\newblock \bibinfo{title}{Superconductivity without insulating states in
  twisted bilayer graphene stabilized by monolayer wse2}.
\newblock \emph{\bibinfo{journal}{Nature}} \textbf{\bibinfo{volume}{583}},
  \bibinfo{pages}{379--384} (\bibinfo{year}{2020}).

\bibitem{park_tunable_2021}
\bibinfo{author}{Park, J.~M.}, \bibinfo{author}{Cao, Y.},
  \bibinfo{author}{Watanabe, K.}, \bibinfo{author}{Taniguchi, T.} \&
  \bibinfo{author}{Jarillo-Herrero, P.}
\newblock \bibinfo{title}{Tunable strongly coupled superconductivity in
  magic-angle twisted trilayer graphene}.
\newblock \emph{\bibinfo{journal}{Nature}} \textbf{\bibinfo{volume}{590}},
  \bibinfo{pages}{249--255} (\bibinfo{year}{2021}).

\bibitem{Sharpe605}
\bibinfo{author}{Sharpe, A.~L.} \emph{et~al.}
\newblock \bibinfo{title}{Emergent ferromagnetism near three-quarters filling
  in twisted bilayer graphene}.
\newblock \emph{\bibinfo{journal}{Science}} \textbf{\bibinfo{volume}{365}},
  \bibinfo{pages}{605--608} (\bibinfo{year}{2019}).

\bibitem{das_symmetry-broken_2021}
\bibinfo{author}{Das, I.} \emph{et~al.}
\newblock \bibinfo{title}{Symmetry-broken {Chern} insulators and {Rashba}-like
  {Landau}-level crossings in magic-angle bilayer graphene}.
\newblock \emph{\bibinfo{journal}{Nature Physics}}
  \textbf{\bibinfo{volume}{17}}, \bibinfo{pages}{710--714}
  (\bibinfo{year}{2021}).

\bibitem{nuckolls_strongly_2020}
\bibinfo{author}{Nuckolls, K.~P.} \emph{et~al.}
\newblock \bibinfo{title}{Strongly correlated {Chern} insulators in magic-angle
  twisted bilayer graphene}.
\newblock \emph{\bibinfo{journal}{Nature}} \textbf{\bibinfo{volume}{588}},
  \bibinfo{pages}{610--615} (\bibinfo{year}{2020}).

\bibitem{choi_correlation-driven_2021}
\bibinfo{author}{Choi, Y.} \emph{et~al.}
\newblock \bibinfo{title}{Correlation-driven topological phases in magic-angle
  twisted bilayer graphene}.
\newblock \emph{\bibinfo{journal}{Nature}} \textbf{\bibinfo{volume}{589}},
  \bibinfo{pages}{536--541} (\bibinfo{year}{2021}).

\bibitem{stepanov2020competing}
\bibinfo{author}{Stepanov, P.} \emph{et~al.}
\newblock \bibinfo{title}{Competing zero-field chern insulators in
  superconducting twisted bilayer graphene}.
\newblock \emph{\bibinfo{journal}{arXiv preprint arXiv:2012.15126}}
  (\bibinfo{year}{2020}).

\bibitem{Andrew2021}
\bibinfo{author}{Pierce, A.~T.} \emph{et~al.}
\newblock \bibinfo{title}{Unconventional sequence of correlated chern
  insulators in magic-angle twisted bilayer graphene}.
\newblock \emph{\bibinfo{journal}{arXiv:2101.04123}}  (\bibinfo{year}{2021}).

\bibitem{Serlin900}
\bibinfo{author}{Serlin, M.} \emph{et~al.}
\newblock \bibinfo{title}{Intrinsic quantized anomalous hall effect in a
  moir{\'e} heterostructure}.
\newblock \emph{\bibinfo{journal}{Science}} \textbf{\bibinfo{volume}{367}},
  \bibinfo{pages}{900--903} (\bibinfo{year}{2020}).

\bibitem{jiang2019charge}
\bibinfo{author}{Jiang, Y.} \emph{et~al.}
\newblock \bibinfo{title}{Charge order and broken rotational symmetry in
  magic-angle twisted bilayer graphene}.
\newblock \emph{\bibinfo{journal}{Nature}} \textbf{\bibinfo{volume}{573}},
  \bibinfo{pages}{91--95} (\bibinfo{year}{2019}).

\bibitem{cao_nematicity_2021}
\bibinfo{author}{Cao, Y.} \emph{et~al.}
\newblock \bibinfo{title}{Nematicity and competing orders in superconducting
  magic-angle graphene}.
\newblock \emph{\bibinfo{journal}{Science}} \textbf{\bibinfo{volume}{372}},
  \bibinfo{pages}{264--271} (\bibinfo{year}{2021}).

\bibitem{rozen2020entropic}
\bibinfo{author}{Rozen, A.} \emph{et~al.}
\newblock \bibinfo{title}{Entropic evidence for a {Pomeranchuk} effect in
  magic-angle graphene}.
\newblock \emph{\bibinfo{journal}{Nature}} \textbf{\bibinfo{volume}{592}},
  \bibinfo{pages}{214--219} (\bibinfo{year}{2021}).

\bibitem{saito2021isospin}
\bibinfo{author}{Saito, Y.} \emph{et~al.}
\newblock \bibinfo{title}{Isospin pomeranchuk effect in twisted bilayer
  graphene}.
\newblock \emph{\bibinfo{journal}{Nature}} \textbf{\bibinfo{volume}{592}},
  \bibinfo{pages}{220--224} (\bibinfo{year}{2021}).

\bibitem{choi2019electronic}
\bibinfo{author}{Choi, Y.} \emph{et~al.}
\newblock \bibinfo{title}{Electronic correlations in twisted bilayer graphene
  near the magic angle}.
\newblock \emph{\bibinfo{journal}{Nature Physics}}
  \textbf{\bibinfo{volume}{15}}, \bibinfo{pages}{1174--1180}
  (\bibinfo{year}{2019}).

\bibitem{xie2019spectroscopic}
\bibinfo{author}{Xie, Y.} \emph{et~al.}
\newblock \bibinfo{title}{Spectroscopic signatures of many-body correlations in
  magic-angle twisted bilayer graphene}.
\newblock \emph{\bibinfo{journal}{Nature}} \textbf{\bibinfo{volume}{572}},
  \bibinfo{pages}{101--105} (\bibinfo{year}{2019}).

\bibitem{kerelsky2019maximized}
\bibinfo{author}{Kerelsky, A.} \emph{et~al.}
\newblock \bibinfo{title}{Maximized electron interactions at the magic angle in
  twisted bilayer graphene}.
\newblock \emph{\bibinfo{journal}{Nature}} \textbf{\bibinfo{volume}{572}},
  \bibinfo{pages}{95--100} (\bibinfo{year}{2019}).

\bibitem{wong2020cascade}
\bibinfo{author}{Wong, D.} \emph{et~al.}
\newblock \bibinfo{title}{Cascade of electronic transitions in magic-angle
  twisted bilayer graphene}.
\newblock \emph{\bibinfo{journal}{Nature}} \textbf{\bibinfo{volume}{582}},
  \bibinfo{pages}{198--202} (\bibinfo{year}{2020}).

\bibitem{zondiner2020cascade}
\bibinfo{author}{Zondiner, U.} \emph{et~al.}
\newblock \bibinfo{title}{Cascade of phase transitions and dirac revivals in
  magic-angle graphene}.
\newblock \emph{\bibinfo{journal}{Nature}} \textbf{\bibinfo{volume}{582}},
  \bibinfo{pages}{203--208} (\bibinfo{year}{2020}).

\bibitem{xu_vortex-like_2000}
\bibinfo{author}{Xu, Z.~A.}, \bibinfo{author}{Ong, N.~P.},
  \bibinfo{author}{Wang, Y.}, \bibinfo{author}{Kakeshita, T.} \&
  \bibinfo{author}{Uchida, S.}
\newblock \bibinfo{title}{Vortex-like excitations and the onset of
  superconducting phase fluctuation in underdoped {La2}-{xSrxCuO4}}.
\newblock \emph{\bibinfo{journal}{Nature}} \textbf{\bibinfo{volume}{406}},
  \bibinfo{pages}{486--488} (\bibinfo{year}{2000}).

\bibitem{zuev2009thermoelectric}
\bibinfo{author}{Zuev, Y.~M.}, \bibinfo{author}{Chang, W.} \&
  \bibinfo{author}{Kim, P.}
\newblock \bibinfo{title}{Thermoelectric and magnetothermoelectric transport
  measurements of graphene}.
\newblock \emph{\bibinfo{journal}{Phys. Rev. Lett.}}
  \textbf{\bibinfo{volume}{102}}, \bibinfo{pages}{096807}
  (\bibinfo{year}{2009}).

\bibitem{PhysRevB.80.081413}
\bibinfo{author}{Checkelsky, J.~G.} \& \bibinfo{author}{Ong, N.~P.}
\newblock \bibinfo{title}{Thermopower and nernst effect in graphene in a
  magnetic field}.
\newblock \emph{\bibinfo{journal}{Phys. Rev. B}} \textbf{\bibinfo{volume}{80}},
  \bibinfo{pages}{081413} (\bibinfo{year}{2009}).

\bibitem{nam2010thermoelectric}
\bibinfo{author}{Nam, S.-G.}, \bibinfo{author}{Ki, D.-K.} \&
  \bibinfo{author}{Lee, H.-J.}
\newblock \bibinfo{title}{Thermoelectric transport of massive dirac fermions in
  bilayer graphene}.
\newblock \emph{\bibinfo{journal}{Phys. Rev. B}} \textbf{\bibinfo{volume}{82}},
  \bibinfo{pages}{245416} (\bibinfo{year}{2010}).

\bibitem{wang2011enhanced}
\bibinfo{author}{Wang, C.-R.} \emph{et~al.}
\newblock \bibinfo{title}{Enhanced thermoelectric power in dual-gated bilayer
  graphene}.
\newblock \emph{\bibinfo{journal}{Phys. Rev. Lett.}}
  \textbf{\bibinfo{volume}{107}}, \bibinfo{pages}{186602}
  (\bibinfo{year}{2011}).

\bibitem{duan2016high}
\bibinfo{author}{Duan, J.} \emph{et~al.}
\newblock \bibinfo{title}{High thermoelectricpower factor in graphene/{hBN}
  devices}.
\newblock \emph{\bibinfo{journal}{Proceedings of the National Academy of
  Sciences}} \textbf{\bibinfo{volume}{113}}, \bibinfo{pages}{14272--14276}
  (\bibinfo{year}{2016}).

\bibitem{ghahari2016enhanced}
\bibinfo{author}{Ghahari, F.} \emph{et~al.}
\newblock \bibinfo{title}{Enhanced thermoelectric power in graphene: Violation
  of the mott relation by inelastic scattering}.
\newblock \emph{\bibinfo{journal}{Phys. Rev. Lett.}}
  \textbf{\bibinfo{volume}{116}}, \bibinfo{pages}{136802}
  (\bibinfo{year}{2016}).

\bibitem{mahapatra2020misorientation}
\bibinfo{author}{Mahapatra, P.~S.} \emph{et~al.}
\newblock \bibinfo{title}{Misorientation-controlled cross-plane
  thermoelectricity in twisted bilayer graphene}.
\newblock \emph{\bibinfo{journal}{Phys. Rev. Lett.}}
  \textbf{\bibinfo{volume}{125}}, \bibinfo{pages}{226802}
  (\bibinfo{year}{2020}).

\bibitem{ghawri2020excess}
\bibinfo{author}{Ghawri, B.} \emph{et~al.}
\newblock \bibinfo{title}{Excess entropy and breakdown of semiclassical
  description of thermoelectricity in twisted bilayer graphene close to half
  filling}.
\newblock \emph{\bibinfo{journal}{arXiv preprint arXiv:2004.12356}}
  (\bibinfo{year}{2020}).

\bibitem{Srivastaveaaw5798}
\bibinfo{author}{Srivastav, S.~K.} \emph{et~al.}
\newblock \bibinfo{title}{Universal quantized thermal conductance in graphene}.
\newblock \emph{\bibinfo{journal}{Science Advances}}
  \textbf{\bibinfo{volume}{5}} (\bibinfo{year}{2019}).

\bibitem{fong2012ultrasensitive}
\bibinfo{author}{Fong, K.~C.} \& \bibinfo{author}{Schwab, K.}
\newblock \bibinfo{title}{Ultrasensitive and wide-bandwidth thermal
  measurements of graphene at low temperatures}.
\newblock \emph{\bibinfo{journal}{Phys. Rev. X}} \textbf{\bibinfo{volume}{2}},
  \bibinfo{pages}{031006} (\bibinfo{year}{2012}).

\bibitem{crossno2016observation}
\bibinfo{author}{Crossno, J.} \emph{et~al.}
\newblock \bibinfo{title}{Observation of the dirac fluid and the breakdown of
  the wiedemann-franz law in graphene}.
\newblock \emph{\bibinfo{journal}{Science}} \textbf{\bibinfo{volume}{351}},
  \bibinfo{pages}{1058--1061} (\bibinfo{year}{2016}).

\bibitem{betz2013supercollision}
\bibinfo{author}{Betz, A.~C.} \emph{et~al.}
\newblock \bibinfo{title}{Supercollision cooling in undoped graphene}.
\newblock \emph{\bibinfo{journal}{Nature Physics}}
  \textbf{\bibinfo{volume}{9}}, \bibinfo{pages}{109--112}
  (\bibinfo{year}{2013}).

\bibitem{PhysRev.181.1336}
\bibinfo{author}{Cutler, M.} \& \bibinfo{author}{Mott, N.~F.}
\newblock \bibinfo{title}{Observation of anderson localization in an electron
  gas}.
\newblock \emph{\bibinfo{journal}{Phys. Rev.}} \textbf{\bibinfo{volume}{181}},
  \bibinfo{pages}{1336--1340} (\bibinfo{year}{1969}).

\bibitem{PhysRevLett.126.216803}
\bibinfo{author}{Srivastav, S.~K.} \emph{et~al.}
\newblock \bibinfo{title}{Vanishing thermal equilibration for hole-conjugate
  fractional quantum hall states in graphene}.
\newblock \emph{\bibinfo{journal}{Phys. Rev. Lett.}}
  \textbf{\bibinfo{volume}{126}}, \bibinfo{pages}{216803}
  (\bibinfo{year}{2021}).

\bibitem{timusk_pseudogap_1999}
\bibinfo{author}{Timusk, T.} \& \bibinfo{author}{Statt, B.}
\newblock \bibinfo{title}{The pseudogap in high-temperature superconductors: an
  experimental survey}.
\newblock \emph{\bibinfo{journal}{Reports on Progress in Physics}}
  \textbf{\bibinfo{volume}{62}}, \bibinfo{pages}{61} (\bibinfo{year}{1999}).

\bibitem{howson_anomalous_1989}
\bibinfo{author}{Howson, M.~A.} \emph{et~al.}
\newblock \bibinfo{title}{An anomalous peak in the thermopower of
  {Y1Ba2Cu3O7}-\${\textbackslash}delta\$ crystals}.
\newblock \emph{\bibinfo{journal}{Journal of Physics: Condensed Matter}}
  \textbf{\bibinfo{volume}{1}}, \bibinfo{pages}{3865--3865}
  (\bibinfo{year}{1989}).

\bibitem{lu_fluctuation_1995}
\bibinfo{author}{Lu, Y.} \& \bibinfo{author}{Patton, B.~R.}
\newblock \bibinfo{title}{Fluctuation thermopower above the superconducting
  transition temperature}.
\newblock \emph{\bibinfo{journal}{Journal of Physics: Condensed Matter}}
  \textbf{\bibinfo{volume}{7}}, \bibinfo{pages}{9247--9254}
  (\bibinfo{year}{1995}).

\bibitem{uri_mapping_2020}
\bibinfo{author}{Uri, A.} \emph{et~al.}
\newblock \bibinfo{title}{Mapping the twist-angle disorder and {Landau} levels
  in magic-angle graphene}.
\newblock \emph{\bibinfo{journal}{Nature}} \textbf{\bibinfo{volume}{581}},
  \bibinfo{pages}{47--52} (\bibinfo{year}{2020}).

\bibitem{PhysRevX.8.031089}
\bibinfo{author}{Po, H.~C.}, \bibinfo{author}{Zou, L.},
  \bibinfo{author}{Vishwanath, A.} \& \bibinfo{author}{Senthil, T.}
\newblock \bibinfo{title}{Origin of mott insulating behavior and
  superconductivity in twisted bilayer graphene}.
\newblock \emph{\bibinfo{journal}{Phys. Rev. X}} \textbf{\bibinfo{volume}{8}},
  \bibinfo{pages}{031089} (\bibinfo{year}{2018}).

\bibitem{PhysRevX.10.031034}
\bibinfo{author}{Bultinck, N.} \emph{et~al.}
\newblock \bibinfo{title}{Ground state and hidden symmetry of magic-angle
  graphene at even integer filling}.
\newblock \emph{\bibinfo{journal}{Phys. Rev. X}} \textbf{\bibinfo{volume}{10}},
  \bibinfo{pages}{031034} (\bibinfo{year}{2020}).

\bibitem{shavit2021theory}
\bibinfo{author}{Shavit, G.}, \bibinfo{author}{Berg, E.},
  \bibinfo{author}{Stern, A.} \& \bibinfo{author}{Oreg, Y.}
\newblock \bibinfo{title}{Theory of correlated insulators and superconductivity
  in twisted bilayer graphene}.
\newblock \emph{\bibinfo{journal}{arXiv:2107.08486}}  (\bibinfo{year}{2021}).

\bibitem{PhysRevX.8.031087}
\bibinfo{author}{Koshino, M.} \emph{et~al.}
\newblock \bibinfo{title}{Maximally localized wannier orbitals and the extended
  hubbard model for twisted bilayer graphene}.
\newblock \emph{\bibinfo{journal}{Phys. Rev. X}} \textbf{\bibinfo{volume}{8}},
  \bibinfo{pages}{031087} (\bibinfo{year}{2018}).

\bibitem{PhysRevB.85.195458}
\bibinfo{author}{Moon, P.} \& \bibinfo{author}{Koshino, M.}
\newblock \bibinfo{title}{Energy spectrum and quantum hall effect in twisted
  bilayer graphene}.
\newblock \emph{\bibinfo{journal}{Phys. Rev. B}} \textbf{\bibinfo{volume}{85}},
  \bibinfo{pages}{195458} (\bibinfo{year}{2012}).

\bibitem{PhysRevLett.124.076801}
\bibinfo{author}{Cao, Y.} \emph{et~al.}
\newblock \bibinfo{title}{Strange metal in magic-angle graphene with near
  planckian dissipation}.
\newblock \emph{\bibinfo{journal}{Phys. Rev. Lett.}}
  \textbf{\bibinfo{volume}{124}}, \bibinfo{pages}{076801}
  (\bibinfo{year}{2020}).

\end{thebibliography}

\pagebreak
\section{Methods}

\subsection{Device fabrication and measurement scheme:}
The devices consist of hBN encapsulated twisted bilayer graphene (tBLG) on a $Si/SiO_2$ substrate. The typical length and width of the devices are $\sim 6\mu m$ and $\sim 2\mu m$, respectively. The usual `tear and stack' technique~\cite{cao2018correlated,cao2018unconventional} is used to fabricate the device and is described in detail in the supplementary information (SI-1). For the resistance measurement, we employ the low-frequency ($\sim 13 Hz$) lock-in technique (SI-3). For the thermopower measurement, an isolated gold line, as shown in Fig.1a, is placed parallel to one side of the tBLG at a separation of $\sim 3\mu m$. Passing a current ($I_{\omega}$) through the heater creates a temperature gradient across the length of the tBLG as depicted by the color gradient (red to blue) in Fig. 1a. As a result, the contact near to the heater will be hotter ($T_h$) compared to the far contact ($T_c$). The temperature of the far contact ($T_c$) is maintained at the bath temperature of the cryo-free dilution fridge by directly anchoring it to the cold finger attached to the mixing chamber plate 
, which we call a cold ground (c.g). To measure $\Delta T$, we have utilized Johnson noise thermometry. As shown in Fig. 1a, the thermometry circuit consists of a LC resonant ($f_r\sim 720kHz$) tank circuit, followed by a cryogenic amplifier (ca). The relay sitting at the mixing chamber plate (Fig. 1a) is used to switch between the thermoelectric voltage and temperature measurement.

\subsection{Activation gaps, band-width and superconducting transition temperature of MtBLG:}
The value of the resistance at the full filling ($\nu=\pm 4$) continuously decreases with increasing temperature up to much higher T $\sim 100K$. 
On the other hand, the value of the resistance at $\nu=0$ and $2$ decreases with increasing temperature up to $\sim 10 K$ and then increases linearly, showing metallic nature (SI-10). These observations are consistent with earlier reports for MtBLG~\cite{cao2018correlated,yankowitz2019tuning,wu2021chern,saito2020independent,lu2019superconductors,cao2018unconventional,arora2020superconductivity,park_tunable_2021}. 
The gap ($\Delta$) determined from the activated plot for  $\nu=0$, $\nu=2$ and $\pm4$ are, respectively, $\sim$ $0.05meV$, $0.25meV$, $11.5 meV$ and  $9.25meV$ as shown in the SI-10 and SI-11. Furthermore, it can be seen (SI-10) that there is a crossover from metallic nature to insulating one at a higher temperature due to interband excitation of the carriers between the flat and dispersive bands. From the crossover temperature, $\sim 150K$ around the Dirac point, one can estimate the bandwidth ($2W$) and found to be of the order of $\sim 10 meV$ for the MtBLG. In Fig. 1e and 1f, we have shown the differential resistance versus bias current with temperature and perpendicular magnetic field around the superconducting dome. In order to extract the transition temperature at a giving filling, we compare the experimental data with the theoretically generated critical current versus temperature using BCS theory, $I_{c}(T) = I_{c}(0)(1-T/T_{c})^{2}$, where the $I_{c}(0)$ is the experimentally measured critical current at $T = 20mK$, and vary the $T_c$ such that the theoretically generated $I_{c}(T)$ traces the experimentally measured $I_{c}$ in Fig. 1e. This was repeated for other carrier densities and shown in the SI-12. The extracted the critical temperature was found to be $\sim 500mK$ at $\nu \sim -2.5$ (SI-12). The measured value of $T_c$ and critical field ($B_{c} \sim 100mT$) of our device matches reasonably well with the available data for MtBLG~\cite{cao2018unconventional,yankowitz2019tuning,lu2019superconductors,arora2020superconductivity,saito2020independent,park_tunable_2021}. 
Note that the Fraunhofer-like pattern in Fig. 1f can be explained by the interference between percolating superconducting paths separated by the normal islands, which is generic feature in MtBLG due to twist-angle in-homogeneity~\cite{zondiner2020cascade,uri_mapping_2020,park_tunable_2021}. These patterns further establish the existence of the superconductivity in our device though the measurement was carried out in two-probe geometry.

\subsection{Theory:} 
As discussed in detail in SI-16, we compute the thermopower and resistivity as a function of filling and temperature for the model of Ref.\cite{zondiner2020cascade} using Hartree-Fock (HF) approximation. The model consists of four spin-valley flavors, interacting with local Coulomb interaction $U$. For the results reported in the main text, we have taken $U=1.2 W$, where $W$ is the band width of the conduction (valence) band. The HF self-consistency equations depend on the non-interacting DOS of the moire' flat bands. We use various non-interacting DOSs, e.g. DOSs obtained from the continuum Bistritzer-MacDonal model~\cite{Bistritzer12233} in Ref.~\cite{zondiner2020cascade} and the DOS generated from the continuum model of Ref.~\cite{PhysRevX.8.031087}, which includes lattice relaxation effects. The self-consistent HF DOS is then used to compute thermopower and resistivity via Kubo formulae neglecting vertex corrections. We assume a constant band velocity and use a small impurity scattering rate $\Gamma_0=0.001W$ in the Kubo formulae.

\section{Acknowledgements}
A.D. thanks the Department of Science and Technology (DST), India for financial support (DSTO-2051), the MHRD, Government of India under STARS research funding (STARS/APR2019/PS/156/FS), and also acknowledges the Swarnajayanti Fellowship of the DST/SJF/PSA-03/2018-19. K.W. and T.T. acknowledge support from the Elemental Strategy Initiative conducted by the MEXT, Japan and the CREST (JPMJCR15F3), JST.

\section{Author contributions}
S.C., A.K.P. and U.R. contributed to device fabrication. A.G. and A.K.P. contributed to data acquisition and analysis. R.D. contributed in initial measurements. A.D. contributed in conceiving the idea and designing the experiment, data interpretation and analysis. K.W and T.T synthesized the hBN single crystals. A.P., A.A., S.M. and S.B. contributed in development of theory, data interpretation, and all the authors contributed in writing the manuscript.

\thispagestyle{empty}
\mbox{}



\section*{Supplementary material}

# SI: Interaction driven giant thermopower in magic-angle twisted bilayer-graphene

Arup Kumar Paul[1], Ayan Ghosh[1], Souvik Chakraborty[1], Ujjal Roy[1], Ranit Dutta[1], Kenji Watanabe[2], Takashi Taniguchi[2], Animesh Panda[1], Adhip Agarwala[3], Subroto Mukerjee[1], Sumilan Banerjee[1] and Anindya Das[1]

[1]Department of Physics, Indian Institute of Science, Bangalore, 560012, India.
[2]National Institute for Materials Science, Namiki 1-1, Ibaraki 305-0044, Japan.
[3]Max Planck Institute for the Physics of Complex Systems, Nöthnitzer Straße 38, 01187 Dresden, Germany.

**SI-1: Device fabrication**

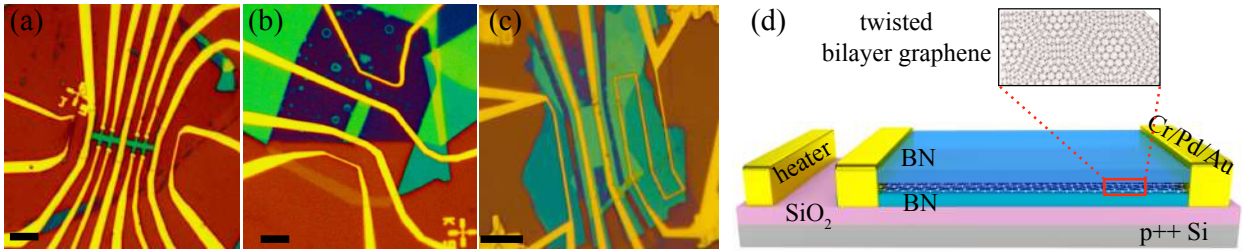

**Supplementary Figure 1:** (a), (b) and (c) Optical images of the measured devices with twist angles $0.26^0$, $1.05^0$ and $1.86^0$, respectively. The scale bars are $5\mu m$. (d) The schematics of the device structure. The devices consist of hexagonal boron nitride (hBN) encapsulated twisted bilayer graphene on $Si/SiO_2$ substrate. The edge contacts with the twisted bilayer devices were made of $Cr(2nm)$, $Pd(10nm)$ and $Au(70nm)$ by thermal evaporation. The $300nm$ wider gold heater line was placed at $\sim 3\mu m$ away from the device.

We use the well-known "tear and stack" [1] method to fabricate the twisted bilayer graphene (tBLG) devices. First, we exfoliate hBN ($\sim 10nm - 25nm$) and graphene on separate Si/SiO$_2$ substrates and then identify the suitable flakes under an optical microscope. After identification, we pick up an hBN flake using a transparent PDMS-poly propyl carbonate (PPC) stamp [1]. This PDMS-PPC-hBN stamp is then used to sequentially pick up the teared graphene layers to form the twisted bilayers. To introduce the twist angle between the graphene layers, we first pick up one graphene part with the PDMS-PPC-hBN stamp, then rotate the substrate containing the other half of the graphene flake. During the graphene pick-up process, we maintain the substrate at a constant temperature between $50^o$-$55^oC$. Finally, we pick up another hBN flake to encapsulate the tBLG layers. The entire stack is then deposited on a clean Si/SiO$_2$ substrate at $75^o$ C.

To make the electrical contacts, we spin-coat with 495A4-950A4 PMMA bilayer. However, instead of the usual temperature of $180^oC$, each PMMA layer were baked at $80^oC$ for 15 minutes. This low-temperature

baking is done to avoid thermal relaxation of the twist angle between the two graphene layers. Then the contacts, including the heater lines, are defined using standard e-beam lithography, with the dose optimized for low baking of PMMA. After developing, the contacts are etched with $CHF_3$-$O_2$ plasma followed by thermal deposition of $Cr(2nm)$, $Pd(10nm)$ and $Au(70nm)$. We have kept the width of source and drain contacts on tBLG $1\mu m$, while the heater lines $\sim 200-300nm$. The heater lines are separated by a $3\mu m$ gap from the source and drain contacts.

## SI-2: Twist angle determination

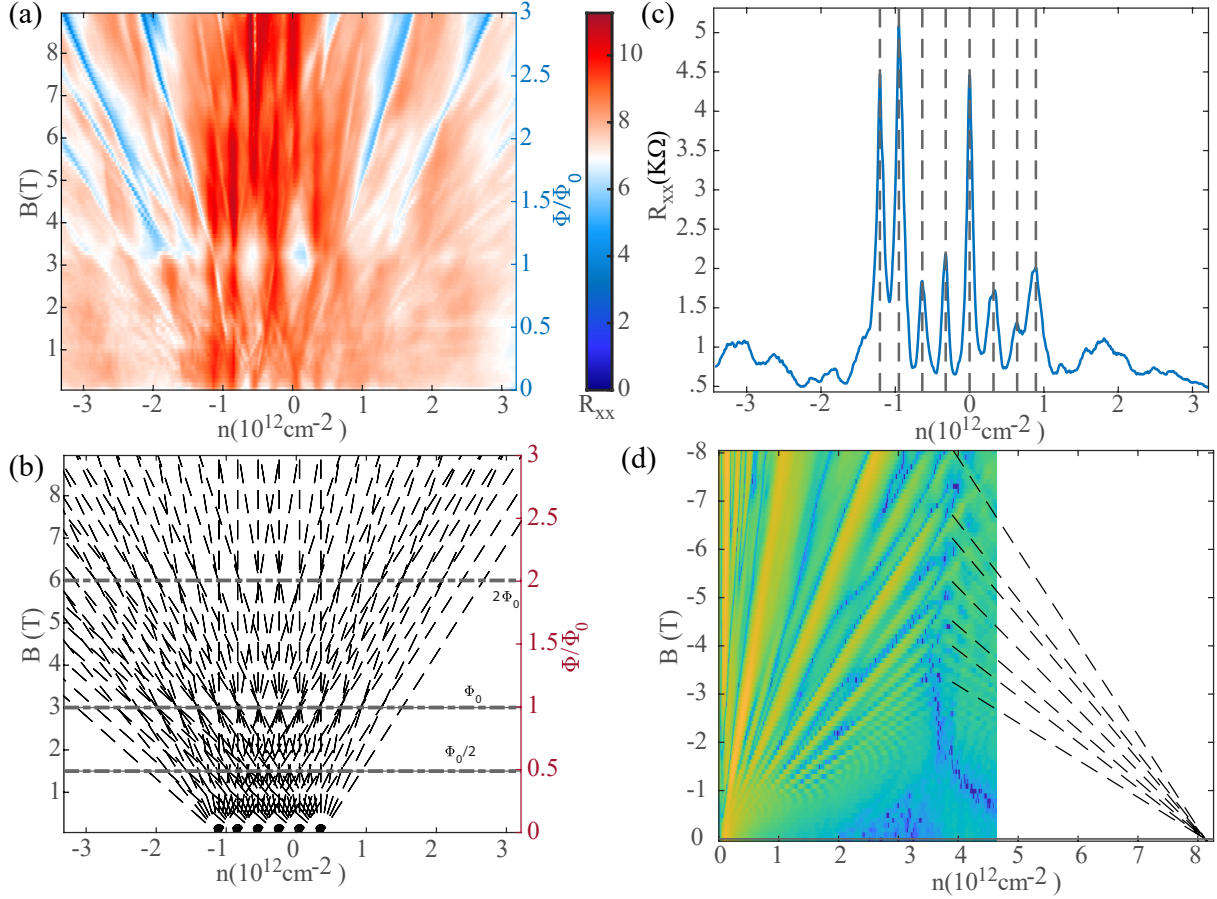

**Supplementary Figure 2: Fan diagrams for tBLG devices:** (a) Fan diagram for ($\sim 0.26^0$) twist-angle device. The plot shows multiple Landau level (LL) structure originating from different densities. (b) Corresponding theoretically generated Diophantine lines shows LL crossing with increasing magnetic fields. (c) Resistance versus density (n) plot for $\sim 0.26^0$ twist-angle device showing multiple resistance peaks. (d) Fan diagram for $\sim 1.86^0$) twist-angle device, with the dashed lines showing extrapolated fan diagram originating from the density corresponds to the band filling. One can clearly see half-filling around $\sim 4 \times 10^{12} cm^{-2}$.

2

**Magic angle sample:** The twist-angle $\theta$ of the MtBLG device is determined from the densities at the full-filling $\nu = \pm 4$ (4 electron per unit cell of the moire superlattice). The area of superlattice unit cell for a twist angle is given by $A \sim \sqrt{3}a^2/2\theta^2$, where $a = 0.246 nm$ is the lattice constant of monolayer graphene. At the full filling, the superlattice density $n_s$ is given by $n_s = 4/A = 8\theta^2/\sqrt{3}a^2$. Using this expression the $\theta$ is determined form the densities at $\nu = \pm 4$ (Figure 1(b) of the main text). In our case for $\nu = \pm 4$ the densities are $n = \pm 2.58 \times 10^{12} cm^{-2}$, which translate into a twist angle of $1.05^0 \pm 0.2^0$ for the magic angle device.

**Marginally twisted sample:** For angles below $0.90^0$, the superlattice gap at full-filling $\pm n_s$ vanishes, and no single particle gap exists, although resistance peaks can be observed at $\pm 2n_s$, caused by Dirac-like band crossings (SI-Fig. 2(c)). Thus, the second resistance peaks after the Dirac point may not correspond to $\pm n_s$ but instead to $\pm 2n_s$. The resistance fan diagram can be used to eliminate this possibility. Each band edge of the superlattice Brillouin zone creates its own Landau level. As can be seen, from the SI- Fig 2(a), the emergence of the landau-fans from $\pm n_s$ marks the full-filling of the flat band. The angle calculated in this method is $0.26^0$. Alternatively, the angle can also be derived from the Hofstadter's Butterfly pattern in the resistance fan. The Landau level crossings are described by Diophantine relation $n/n_s = c\nu\phi/\phi_0 + S$ where $\phi$ is the magnetic flux through a unit cell, $\nu$ is an integer, $\phi_0 = h/e$ is the flux quanta and $S = 0$ labels the main Landau fan, $S = \pm 1$ is the first satellite fan, and so on. Two adjacent landau level crosses when $\phi_0/\phi = q$ or, $B = \phi_0/qA$, $q$ being a half-integer quantity. Thus, the crossing for $q = 1$ should occur at $B = \phi_0/A$ and so on. Thus, knowing the landau level crossings in the magnetic field, the superlattice area, and the twist angle can be calculated. The SI-Fig. 2(b) shows Diophantine relation nicely fits the experimental fan diagram. In this case $q = 1$ corresponds to $B = 3T$ which gives $\theta = 0.25^0$.

**Large angle sample:** For the larger angle sample following the expression of $n_s$, it is evident that the superlattice density is much higher compared to the magic angle sample. Probing such higher densities requires high gate voltage applied to the $SiO_2$ gate which can cause a dielectric breakdown. The transport measurement, as a result, was limited to densities less than full-filling density. Therefore, to accurately determine the full-filling density, the fans appearing from the band edge has to be extrapolated to find out the density from which they emanate. From SI-Fig. 2(d), one can clearly see the changeover slope of the Fan around $n \sim 4 \times 10^{12} cm^{-2}$. Tracing those Fans beyond $n = 4 \times 10^{12} cm^{-2}$ we have extrapolated the dashed lines in SI-Fig 2(d) and can be seen that they merge to a single point corresponds to the full-filling density, $n = \pm 8.12 \times 10^{12} cm^{-2}$. This gives us $\theta = 1.86^0$.

### SI-3: Measurement setup

The schematic of the complete measurement setup is shown in SI-Fig. 3(a). Here the red dashed box indicates the part of the measurement setup located on the cold finger attached to the mixing chamber plate (MC plate). As described in the main text, the sample contact far from the heater line is directly bonded to the cold finger, which serves as a cold ground (c.g). On the other hand, the contact adjacent to the heater denoted as hot contact is used for the resistance, $V_{2\omega}$ as well as for $\Delta T$ measurement. The hot contact has two connections: one going to the left most coaxial cable via a 1M$\Omega$ resistance, and another going to the relay, which is used to switch between the low frequency (resistance and $V_{2\omega}$ measurement) and thermal noise measurement scheme. SI-Fig. 3(b) shows the simplified resistance measurement scheme. Here, the left most coaxial line is used send low frequency (13 Hz) current ($I_{ac}$) through the sample. The high

**Supplementary Figure 3: Setup for the measurement:** (a) Schematic of the measurement setup. (b), (c) and (d) shows simplified form of the resistance, $V_{2\omega}$ and $\Delta T$ measurement schemes, respectively.

resistance in series serves as ballast for this purpose. The voltage drop ($V_S$) across the sample is measured through the right most coaxial line using a SR560 voltage pre-amplifier followed by a lock in Amplifier. Here $R_s$ corresponds to the sample resistance. SI-Fig. 3(c) shows the simplified $V_{2\omega}$ measurement scheme. The middle two coaxial cables are used to send ac ($I_h$) through the heater line at 13 Hz and a $1k\Omega$ resistance connected in series to the heater line acts as the ballast resistance. Similar to the resistance measurement the Seebeck voltage or the second harmonic $V_{2\omega}$ signal generated across the tBLG devices are measured via the rightmost coax cable. The $\Delta T$ is measured by noise thermometry. The SI-Fig. 3(d) shows the simplified schematic of the $\Delta T$ measurement setup. The thermal noise setup consists of an LC resonant circuit ($f_r \sim 720 kHz$), followed by a low noise HEMT cryo-amplifier, placed on the 4K plate (shown as the sky-blue dashed box) of the dilution refrigerator. The noise signal is further amplified by a room temperature amplifier (RTA) and then goes to a spectrum analyzer. To measure the thermal noise generated across the sample the hot contact of the sample is switched to the input of the LC circuit using the relay. For this measurement, dc current is sent through the heater line to create the temperature gradient across the sample.

### SI-4: Temperature gradient along the measurement line

As shown in SI-Fig. 4, the temperature difference (from $100mK - 300K$) between the sample and the SRS560 amplifier can create spurious DC Seebeck voltages along the measurement line. These DC voltages will not affect the thermopower measurement since the measurement is done by ac $2\omega$ technique. However, if proper care is not taken, these additional DC voltages can send significant DC current through the sample.

4

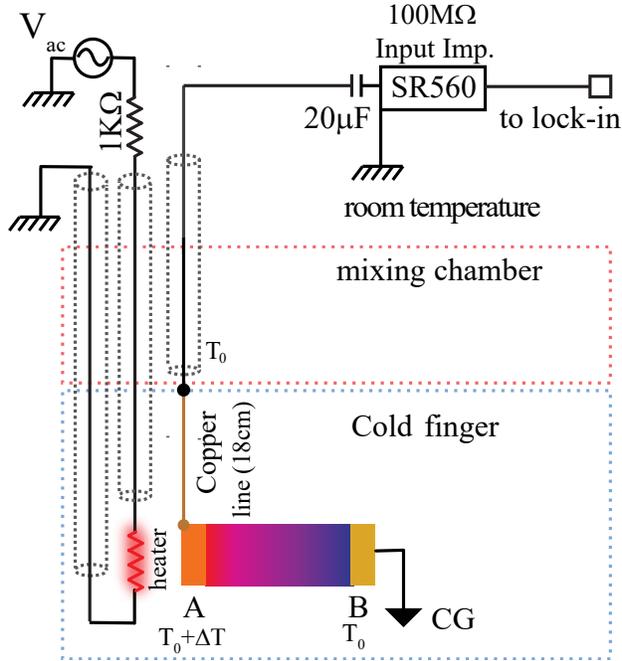

**Supplementary Figure 4: Schematic of the measurement lines:** The different part of the measurement lines are at different temperatures.

This extra DC current can increase the overall sample temperature by Joule heating, thus rendering our $\Delta T$ measurement inaccurate. Also it can destroy phenomena like superconductivity. Thus eliminating any potential DC bias dropped across the device due to constant temperature gradient is critical. SI-Fig. 4 is a simplified schematic emphasizing the different temperature regions across the measurement circuit. Now, if we follow the Seebeck voltage measurement path from sample contact B to the input of the SR560 pre-amplifier (room temperature), it can seen that a constant temperature gradient is always maintained across the entire line. However the generated DC Seebeck voltage primarily drops at the SR560 as it has input impedance of $\sim 100M\Omega$, much higher than the resistances of the sample. Furthermore, we have also added a $20\mu F$ capacitor at the SR560 input, effectively allowing the DC Seebeck voltage to drop across the capacitor only. Moreover, all the coaxial cables going to sample and heater are thermally anchored at every stages of the Dilution fridge. The sample is mounted on a cold finger attached to the mixing chamber. The contact B is directly bonded to the cold finger, and the line from contact A of the sample to the coaxial line is a $18cm$ long copper wire (0.5mm thickness) which is adhered to the cold finger via silver paint for good thermal coupling [2]. When the oscillating current passes through the heater, it oscillates the temperature of hot junction A to a temperature $T_0 + \Delta T$, while the whole cold finger and mixing chamber plate stay at $T_0$. This causes a temperature gradient across the copper line connecting junction A (at temperature $T_0+\Delta T$) to the coaxial cable (at temperature $T_0$) on the mixing chamber plate. However, the resultant Seebeck voltage can be ignored (in the measurement temperature range $\leq 40K$), because Seebeck for copper is of the order of $nVK^{-1}$, which is 3 decades smaller compared to the Seebeck of our devices.

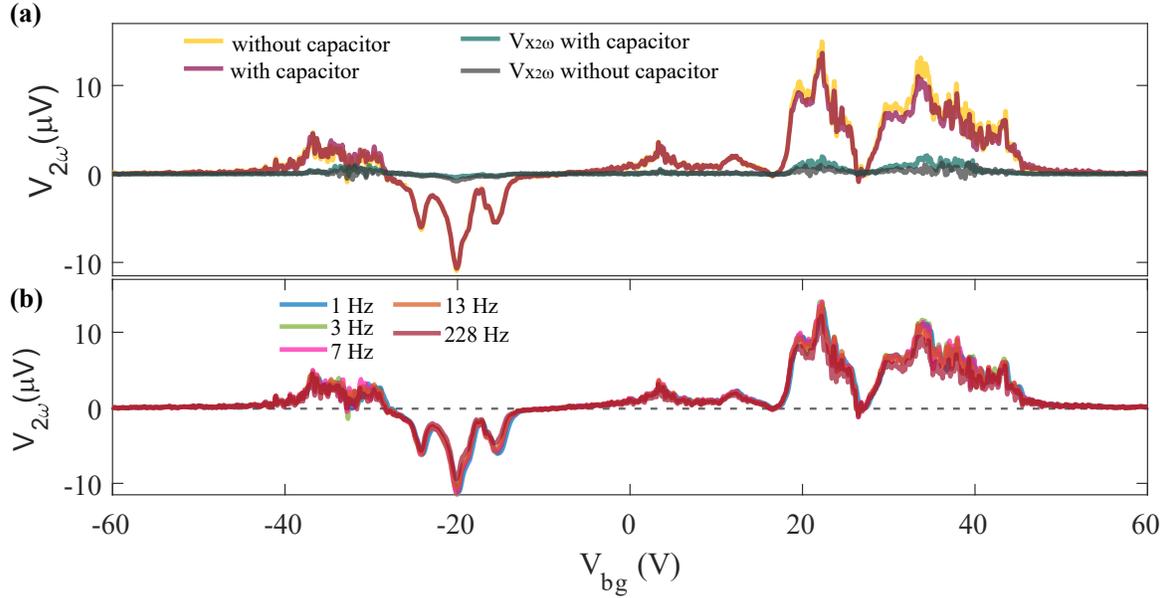

**Supplementary Figure 5: Effect of measurement frequency:** (a) $V_{2\omega}$ versus $V_{bg}$ at $200mK$ with and without capacitor. (b) $V_{2\omega}$ versus $V_{bg}$ at $200mK$ for different frequencies.

### SI-5: Effects of measurement frequency on $V_{2\omega}$ signal

It can be seen from SI-Fig. 5(a) that the measured $V_{2\omega}$ signal remains insensitive of whether the input capacitor to SRS560 is there or not. For the completeness we have shown the 'X' component of the measured signal ($V_{X2\omega}$), which remain almost negligible compared to the 'Y' component $V_{2\omega}$ signal. The $V_{2\omega}$ signal in SI-Fig. 5(a) was measured using the $2\omega$ technique at $13Hz$ for a different thermal cycle than the data presented in the main manuscript. SI-Fig. 5(b) compares the $V_{2\omega}$ for different frequencies without the blocking capacitor at $200mK$. As evident from the figure, frequency does not have any effect on the $V_{2\omega}$ measurement. This rules out any experimental artifact for showing thermopower peak around the superconducting dome (around $V_{bg} \sim -20V$) as described in Ref [3], where it is expected to change drastically with frequency.

### SI-6: Excess thermal noise $S_V = 2k_B \Delta TR$

As mentioned in the manuscript to measure $\Delta T$, we have utilized Johnson noise thermometry [2, 4, 5, 6, 7, 8]. Without any heater current, the entire device remains in equilibrium with the MC plate. Hence, the measured thermal noise by the cryo-amplifier is given by $S_V{}^0 = 4k_B T_c R_s$, where $T_c$ corresponds to the temperature of the MC plate and $R_s$ is the resistance of the device. In the linear regime, the temperature profile across the tBLG as function of distance ($x$) from the hot contact is shown in SI-Fig. 6 when the heater current is non-zero. Here the position $x = 0$, corresponds to the junction of tBLG and hot contact at temperature $T_h = T_c + \Delta T$, with $\Delta T$ being the net temperature difference between the hot and cold contacts.

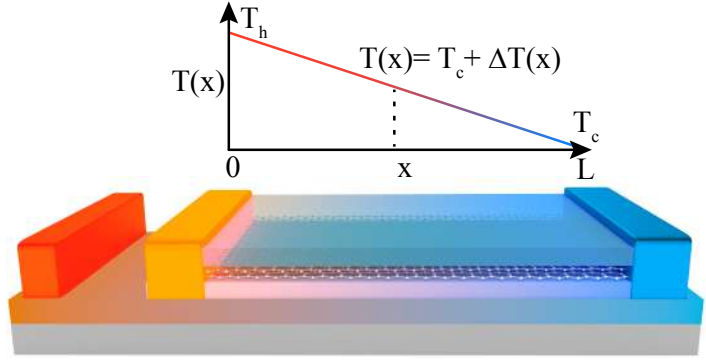

**Supplementary Figure 6:** Temperature gradient ($\Delta T$) across the device in presence of heater current ($I_h$).

The position $x = L$, corresponds to the junction of graphene and cold contact at temperature $T_c$. Thus, at any position on the tBLG the temperature is given by $T(x) = T_c + \Delta T(x)$, where $\Delta T(x) = \Delta T(1-x/L)$. In this scenario The total thermal noise seen by the amplifier is the sum of noise coming from all the infinitesimal parts of the tBLG. Hence, the measured noise ($S_V{}^I$) is given by:

$$S_V{}^I = 4k_B \int_0^L (T_c + \Delta T(x))\rho(x)dx \tag{1}$$

where, $\rho(x)$ corresponds to the resistivity of the TBLG. Considering homogeneous resistance i.e $\rho(x) = R_s/L$, the eqn. 1 can be rewritten as

$$S_V{}^I = 4k_B T_c R_s + 4k_B \rho \int_0^L (\Delta T(x))dx$$
$$= 4k_B T_c R_s + 4k_B \rho \Delta T \int_0^L (1-x/L)dx$$
$$= 4k_B T_c R_s + 2k_B \Delta T R_s$$

Thus, the excess thermal noise due to temperature gradient is given by $S_V = S_V{}^I - S_V{}^0 = 2k_B \Delta T R_s$.

**SI-7: Measuring $\Delta T$**

To measure $\Delta T$, we first measure the thermal noise by varying the heater current from zero. The excess thermal noise as described in the previous section, is determined by subtracting the zero current thermal noise from the total measured noise at any given current. SI-Fig. 7 shows the measured excess thermal noise ($S_V$) as function of square of applied heater current ($I^2$) at different temperatures. As it can be seen the measured $S_V$ is linear with $I^2$, thus again corroborates the linear response regime. The temperature gradient $\Delta T$ is derived from the $S_V$ data using $S_V = 2k_B R \Delta T$ as shown in SI-Fig. 8. The red lines are the linear fit showing linearity of $\Delta T$ with $I^2$.

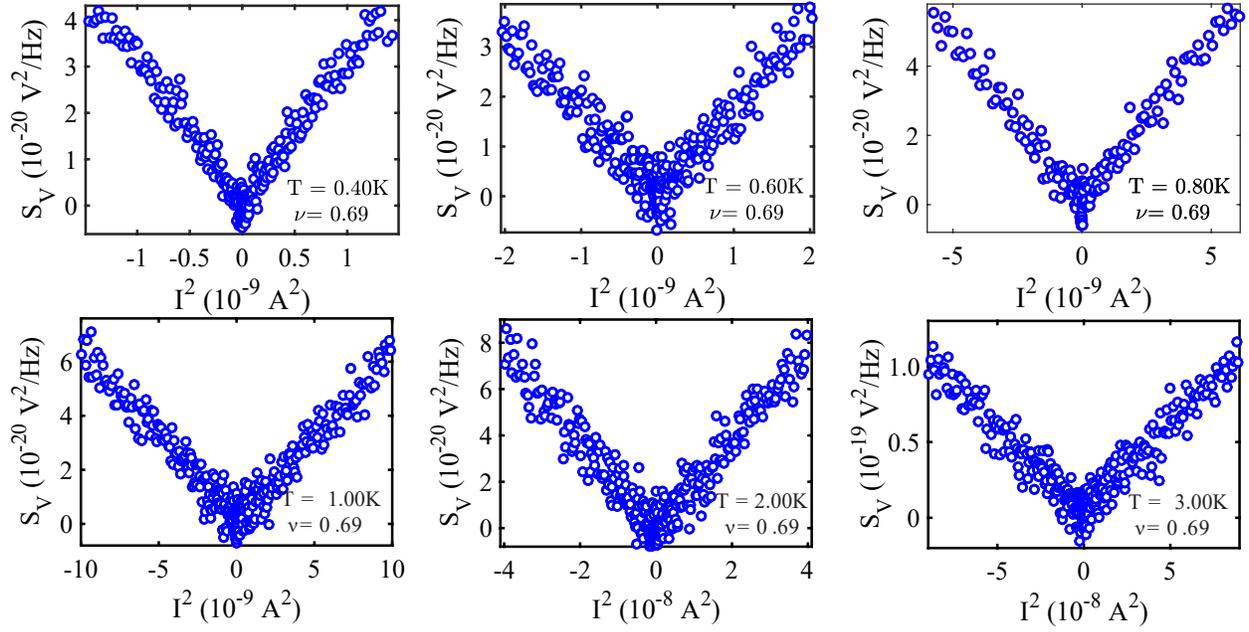

**Supplementary Figure 7:** Thermal noise ($S_V$) as function of square of applied heater current ($I^2$) for different temperatures. As can be seen $S_V$ in linear to $I^2$. The sign indicates the direction of current

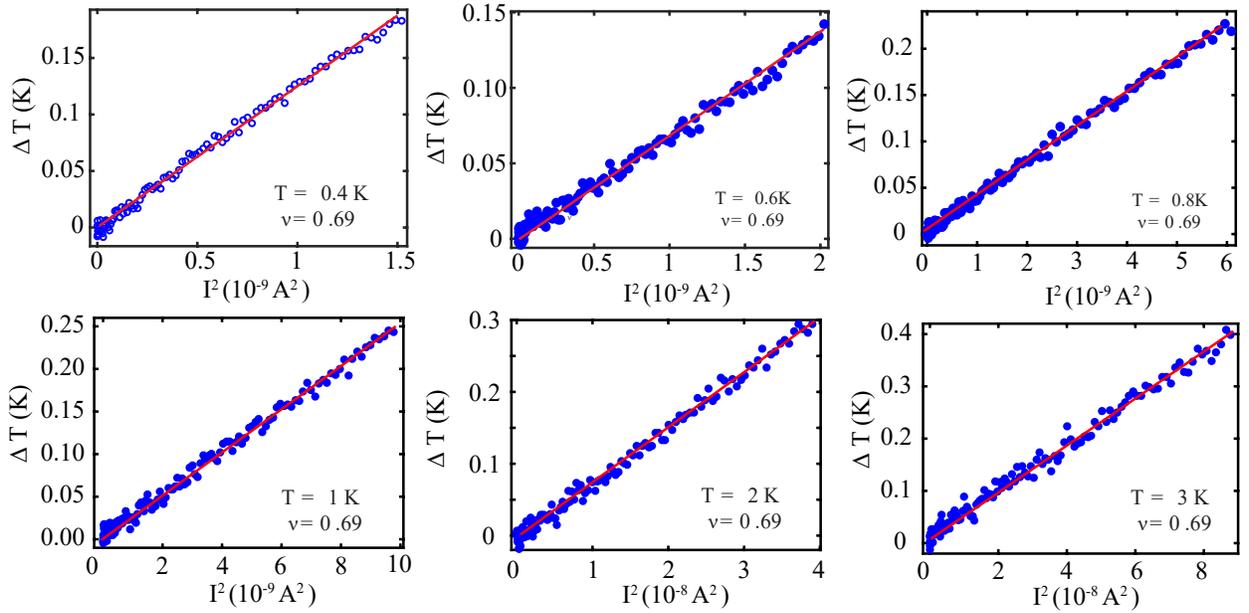

**Supplementary Figure 8:** Measured $\Delta T$ as a function of square of applied heater current ($I^2$) for different temperatures.

8

SI-Fig. 9(a) shows $\Delta T$ versus $I^2$ near Dirac point $\nu = 0.4$ (red) and inside the dispersive band $\nu = 5.9$ (blue) at $T = 10K$. As can be seen the measured $\Delta T$ does not depend on the carrier concentration of the sample. Further, SI-Fig. 9(b) shows $\Delta T$ versus $I^2$ at $\nu = 0.4$ at $T = 1K$ without (red) and with (blue) perpendicular magnetic field. It can be seen that the measured $\Delta T$ does not depend on magnetic field. These observations indicate that $\Delta T$ is determined by the substrate.

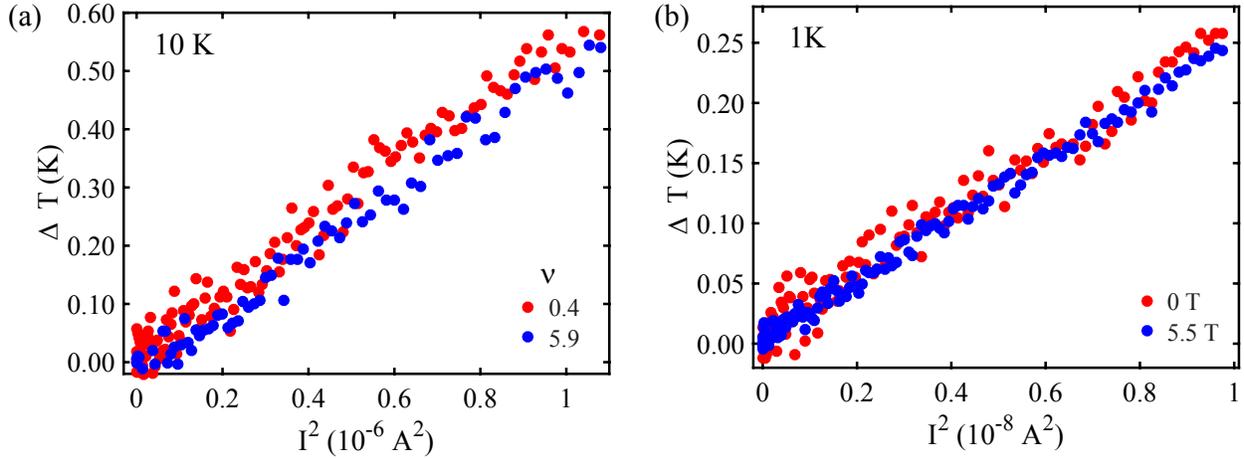

**Supplementary Figure 9:** (a) Measured $\Delta T$ versus $I^2$ for different densities $\nu = 0.4$ (red) and $\nu = 5.9$ (blue) at $T = 10K$. (b) $\Delta T$ versus $I^2$ at $\nu = 0.4$ for zero (red) and $B = 5.5T$ (blue) at $T = 1K$.

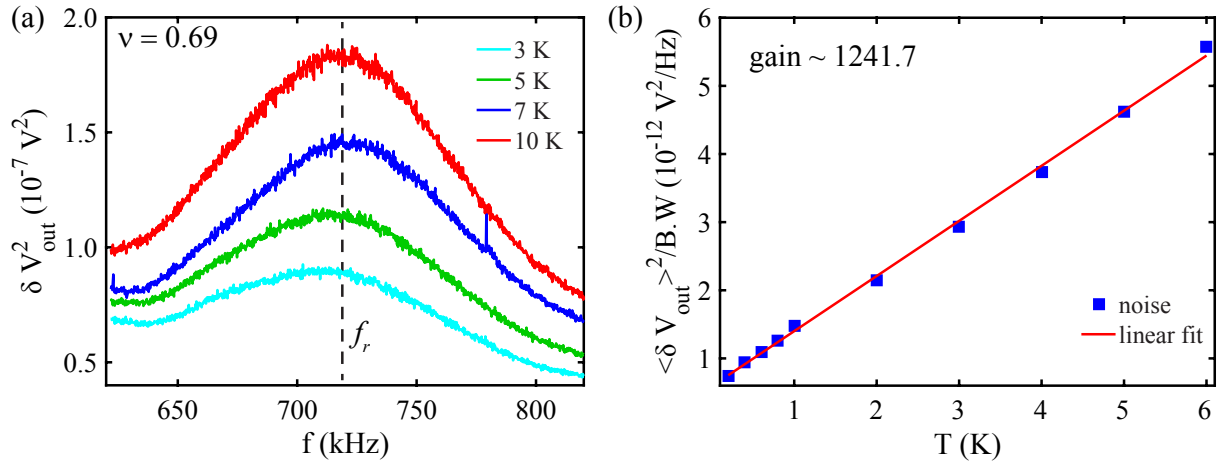

**Supplementary Figure 10:** (a) Square of measured noise voltage output $\delta V_{out}^2$ of the amplifier chain as a function of measurement frequency ($f$) for different temperatures at $\nu = 0.69$. The resonance frequency ($f_r$) is $\sim 720 kHz$. (b) Measured thermal noise ($<\delta V_{out}>^2 /B.W$) as function of $T$ at $f_r$ showing linear dependence. Here B.W is the measurement bandwidth. Red line shows linear fit from which the gain ($\sim 1241$) is determined.

## SI-8: Gain calibration of the amplifier

The gain of the cold amplifier-room temperature amplifier chain described in SI. 3 is crucial for the accurate determination of $\Delta T$. We use noise thermometry to measure the gain. For this first, we determine the resonance frequency ($f_r$) of the LC tank circuit at the input of the cold amplifier as the LC circuit acts as a band-pass filter and, at resonance, has unit gain. Supplementary Fig. 10(a) shows the square of r.m.s noise voltage output $\delta V_{out}^2$ of the amplifier chain as a function of measurement frequency ($f$) for different temperatures at $\nu = 0.69$. We have chosen this particular filling as the resistance ($\sim 9.5\ k\Omega$) remains invariant up to $T = 10K$. As can be seen from the figure, all the $\delta V_{out}^2$ vs. ($f$) plots show a maxima around $\sim 720\ kHz$, which corresponds to the resonance frequency. Then $<\delta V_{out}>^2$ is measured at the resonance frequency within a bandwidth (B.W) of $30 kHz$ as a function of temperature. In SI-Fig. 10(b) $<\delta V_{out}>^2 /B.W$ is plotted against temperature. The plot shows the linear temperature dependence, which is expected as for any resistance R, the thermal noise is given by $S_V = 4k_B T R = <\delta V_{out}>^2 /(B.W \times A^2)$, where $K_B$ is the Boltzmann constant and A is the gain. The gain is determined from the slope of the plot as $A = \sqrt{\frac{slope}{4k_B R}}$ and is found to be around $\sim 1241$.

## SI-9: Linear response regime

The SI-Fig. 11(a) and (b) are the $V_{2\omega}$ versus $V_{bg}$ for different applied heater current, respectively, at temperature 1 K and 20 K. In SI-Fig. 11(c) $V_{2\omega}$ is plotted as function of square of heater current $I_h$ at different fillings for T = 1 K. Similar plot for T = 20 K is plotted in SI-Fig. 11(d). As can be seen form the figures within the maximum applied heater current $V_{2\omega}$ is linear with $I_h^2$, showing that the $V_{2\omega}$ measurements have been performed in the linear response regime.

## SI-10: Resistance versus temperature of MtBLG

The temperature evolution of resistance versus $\nu$ response is shown in SI-Fig. 12(a). As can be seen from the figure for $\nu = \pm 4$ resistance always increases as temperature is reduced, indicating conventional band insulator like nature. However for $|\nu| < 4$, as temperature is reduced, resistance initially increases, then again starts to reduce. The temperature of this transition ($T_n$) depends on the filling and is depicted by the black line on the plot. As can be seen the maximum $T_n$ is at Dirac point and as $\nu$ approaches $\pm 4$, $T_n$ reduces. This behavior suggest that carrier transport inside the flat bands at elevated temperatures is dominated by both intra-band and thermally excited inter-band carriers [9]. As the temperature is reduced thermally excited carrier population reduces. As a result initial increase in resistance is observed. The excitation energy depends on the band gap between the flat and dispersive bands and the relative position of the Fermi level from the band edges. Thus at Dirac point maximum energy is required hence $T_n$ is maximum. On the other hand when $\nu$ approaches full filling the excitation energy is essentially becomes equal to the band gap, hence $T_n$ reduces. Below $T_n$ intra-band carriers starts to dominate the transport. The SI-Fig. 12(b) shows the temperature derivative of resistance ($dR/dT$) versus $\nu$ at $T \sim 50K$. The metallic ($dR/dT > 0$) to insulating ($dR/dT < 0$) switching can be seen in the plot. SI-Fig. 12(c) and 12(d) shows the temperature dependence of R near and at Dirac point $\nu = 0$ and at integer fillings for $T < 100K$. It can be seen from the plots that for $\nu = \pm 4$, R versus T shows insulating dependence while for $\nu = -2$ complete

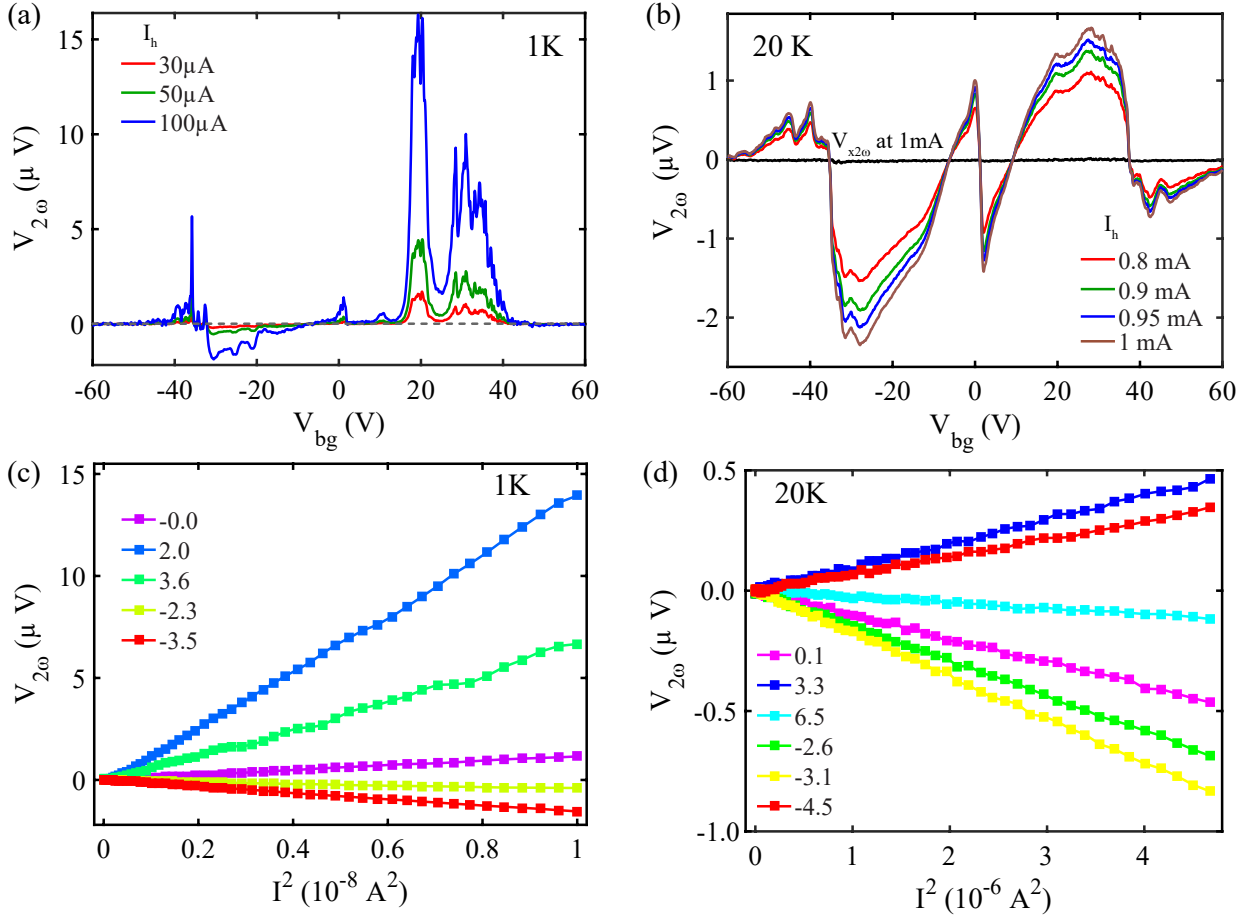

**Supplementary Figure 11:** (a) and (b) $V_{2\omega}$ versus $V_{bg}$ for different applied heater currents, respectively, at temperature 1 K and 20 K. (c) and (d) $V_{2\omega}$ versus $I^2$ for different filling factors, respectively, at temperature 1 K and 20 K. Within the applied current range $V_{2\omega}$ is linear with $I^2$.

metallic nature is observed. At the Dirac point below $20K$ and at $\nu = 2$ below $10K$ insulating behaviour reappears.

### SI-11: Gap determination

From the activated nature of the resistance at integer fillings described in SI-10 we determine the associated gaps using Arrhenius equation. log of Resistance $(lnR)$ as function of inverse of temperature $(T^{-1})$ at zero magnetic field for Dirac point ($\nu = 0$), half filling ($\nu = 2$) and at full fillings $\nu = \pm 4$ are, respectively, shown in SI-Fig. 13(a), 13(b), 13(c) and 13(d). The gap ($\Delta$) determined from the plot for $\nu = 0$, $\nu = 2$ and $\pm 4$ are, respectively, $0.05 meV$, $0.26 meV$, $11.5 meV$ and $9.25 meV$. From the crossover temperature $(T_n)$, $\sim 150K$ around the Dirac point, one can estimate the bandwidth $(2W)$ and found to be of the order of $\sim 10 meV$ for the MtBLG.

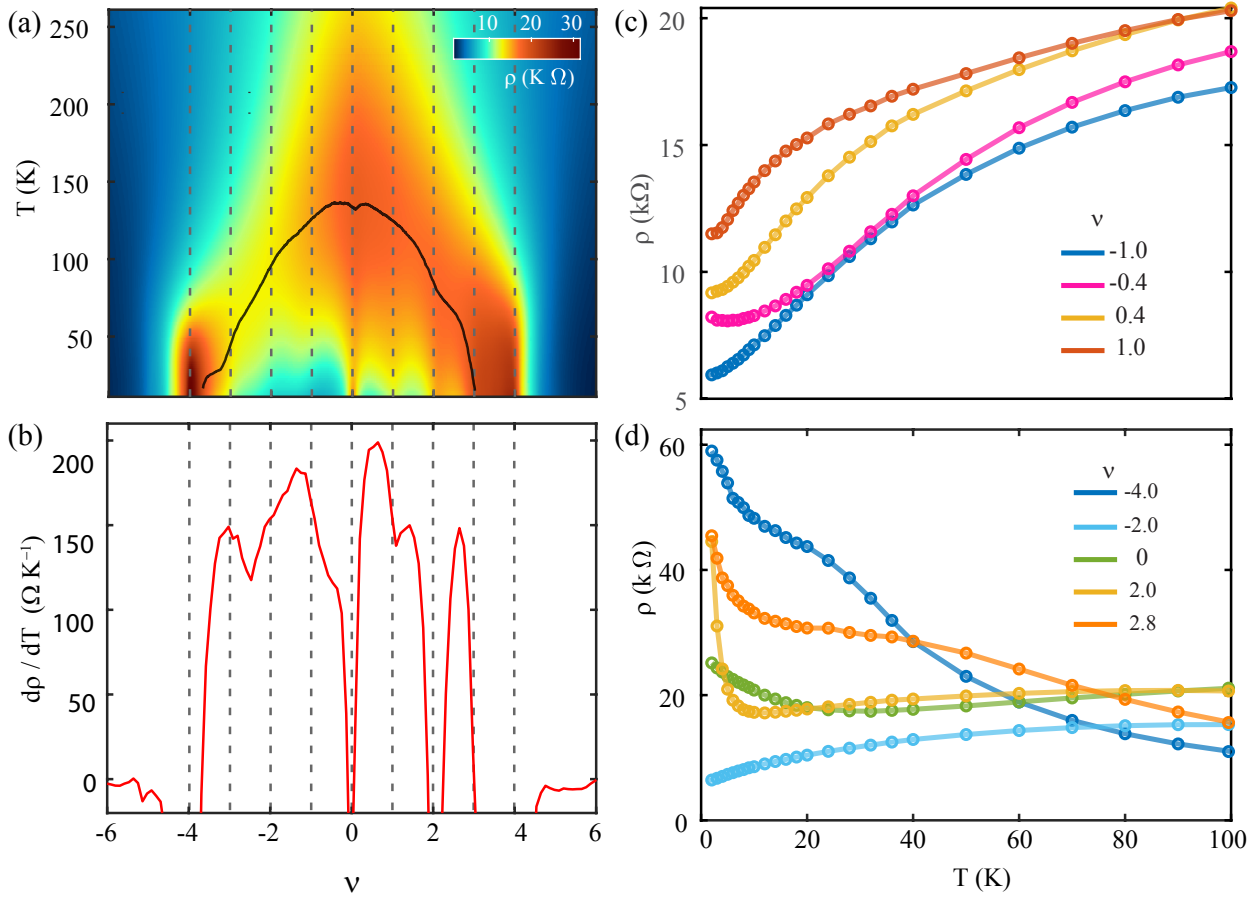

**Supplementary Figure 12:** (a) Temperature evolution of R versus $\nu$ response. (b) $dR/dT$ versus $\nu$ showing metal ($dR/dT > 0$) to insulator ($dR/dT < 0$) switching as function of $\nu$. (c) and (d) R versus T at Dirac point and at integer fillings for T $< 100K$.

**SI-12: Resistance response across the superconducting dome**

SI-Fig. 14 is the summary of temperature and magnetic field dependence of resistance around the superconducting region ($-4 < \nu < -1.5$). The 2D plot in SI-Fig. 14(a) upper panel shows the evolution R versus $\nu$ with increasing temperature. As can be seen the superconducting region is divided into two distinct dome, by a region with lowest critical temperature ($T_C \sim 0.17K$) around $\nu \sim -2.2$. At the superconducting region the resistance saturates to $\sim 1.8k\Omega$, which is due to two-probe measurement scheme. The left side dome has the highest $T_C$ (500mK) at $\nu \sim -2.75$. The lower panel shows $R$ versus $\nu$ response without and with a small magnetic field ($0.1T$) at base temperature $\sim 20mK$. While at zero magnetic field response shows a plateau with a resistance $\sim 1.8k\Omega$, but the response at $0.1T$ shows emergence of an resistance peak at $\nu \sim -2.3$, whose position coincides with the lowest $T_C$ region ($-2.4 < \nu < -2.1$). This is similar to previous observations [1, 10]. In SI-Fig. 14(b), $R$ versus $T$ dependence for different filling factor within $-2.7 < \nu < 4$ are plotted, which shows the switching from superconducting ($\nu = -2.7$) to metallic ($\nu = -3.2$) to eventually insulating ($\nu = -3.6$) states. SI-Fig. 14(c) upper panel shows the magnetic

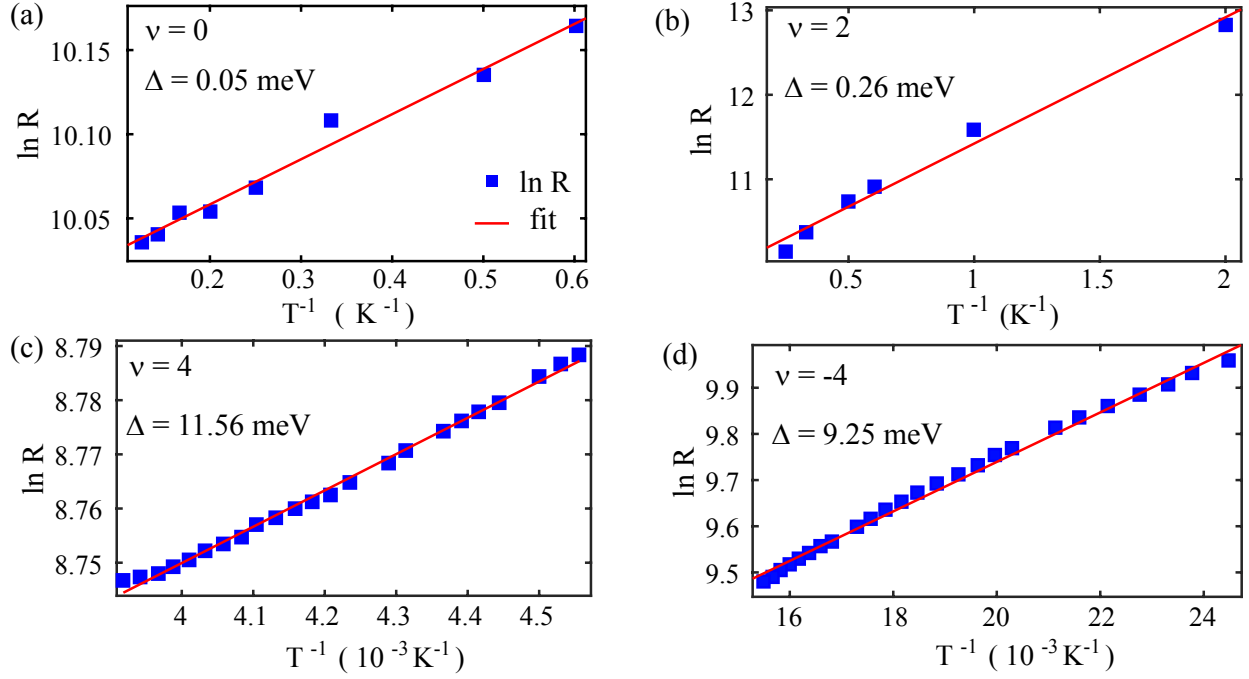

**Supplementary Figure 13:** $lnR$ versus inverse of temperature($T^{-1}$) at zero magnetic field at (a) Dirac point ($\nu = 0$), (b) half filling ($\nu = 2$), (c) and (d), respectively, at full fillings $\nu = 4$ and $\nu = -4$. The gap ($\Delta$) determined from the plot for $\nu = 0$, $\nu = 2$ and $\pm 4$ are $0.05 meV$, $0.26 meV$, $11.5 meV$ and $9.25 meV$.

field dependence of critical current ($I_c$) at $\nu = -2.65$ at base temperature. Here differential resistance ($R_d$) is plotted as function of applied bias current $I_{Bias}$ for different magnetic fields (0 to $100mT$). $I_c$, which corresponds to the sharp transition in $R_d = dV/dI$ shows the usual exponential reduction in nature with increasing magnetic field. The lower panel 2D-plot shows the temperature evolution of $R_d$ vs. $I_{Bias}$ response at the $\nu = 2.7$. In the figure the dome shaped dark region corresponds to the superconducting part and its boundary depicts the $I_c$ vs $T$ dependence. The white dashed line shows the theoretically generated $I_c$ as function of $T$ from the BCS equation $I_c = I_c(0)(1 - (T/T_c)^2)$, where $T_c$ is the critical temperature. As can be seen the BCS equation perfectly captures the observed $I_c$ as function versus $T$ dependence. Similar temperature evolution plots for $\nu = -2, -2.8$ and $-2.98$ are shown, respectively, in SI-Fig. 14(d), 14(e) and 14(f), with the theoretical BCS plot shown by white dashed lines.

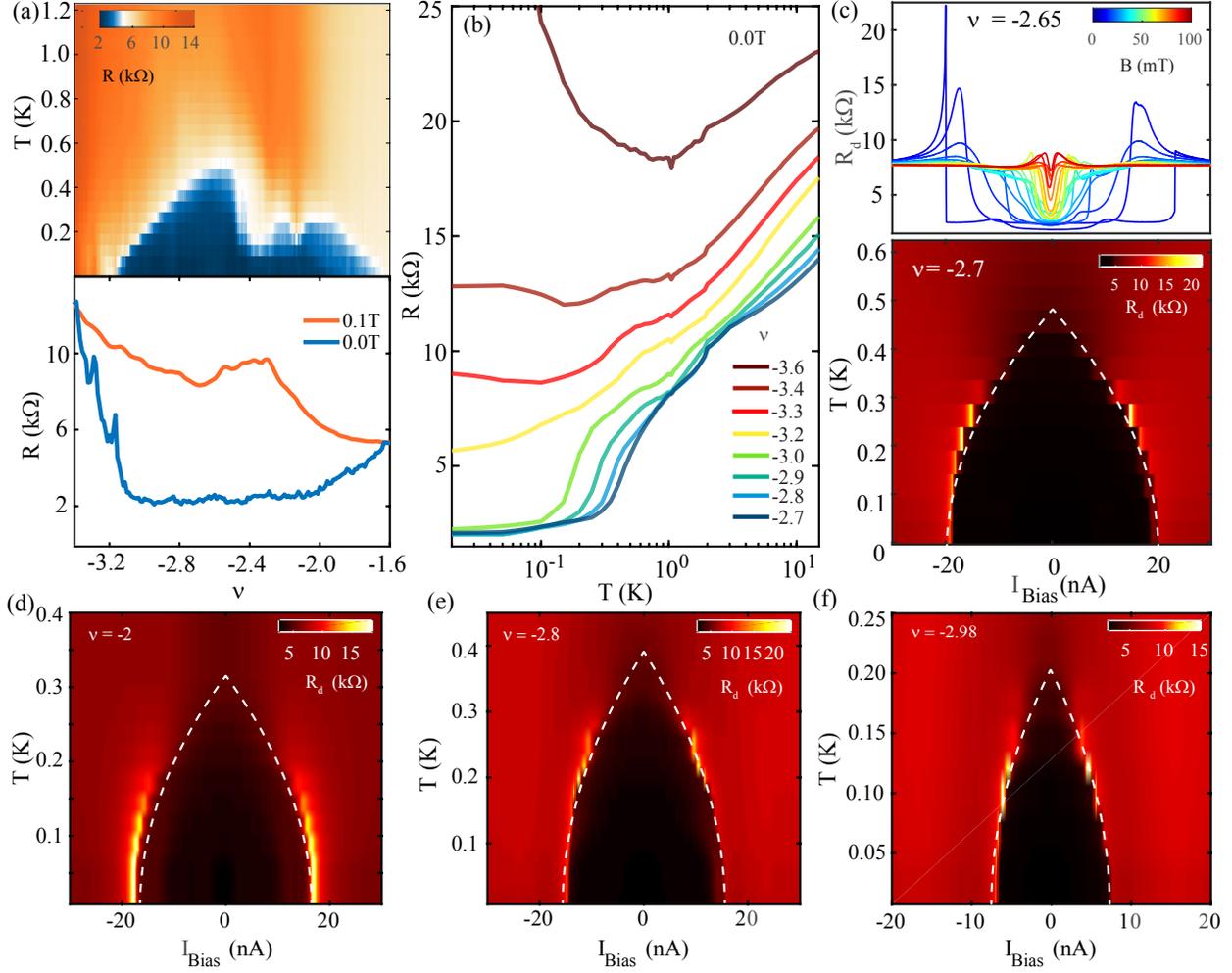

**Supplementary Figure 14:** (a) Upper panel shows temperature evolution of resistance (R) versus $\nu$ around the superconducting region ($-4 < \nu < -1.6$). Lower panel shows $R$ versus $\nu$ without (blue) and with (red) magnetic field (0.1 T). (b) $R$ versus $T$ response for different filling factors within $-4 < \nu < -2.7$. The plot shows transition from superconducting to insulating phase via metallic phase as $\nu$ approaches full filling ($\nu = -4$). (c) Evolution of differential resistance ($R_d = dV/dI$) versus applied current ($I_{Bias}$) with magnetic field (upper panel) at $\nu = -2.65$. Lower panel with temperature at $\nu = -2.7$. The white dashed lines in the lower panel are theoretically generated $I_c$ vs. T dependence from BCS theory. (d),(e) and (f) are similar temperature evolution plot, respectively, for $\nu = -2, -2.8$ and $-2.98$.

**SI-13: Violation of Mott formula**

SI-Fig. 15(a) shows comparison of measured $S$ versus $\nu$ plot (red), with $dR/dn$ versus $\nu$ plot (blue) for the MtBLG. In order to plot them together, here, we have scaled the $dR/dn$ with the measured $S$ of the dispersive band. According to the Mott formula $dR/dn$ term is responsible for sign change observed in thermopower. As can be seen for both $T = 10K$ (upper panel) and $T = 5K$ (lower panel), the $S$ versus

14

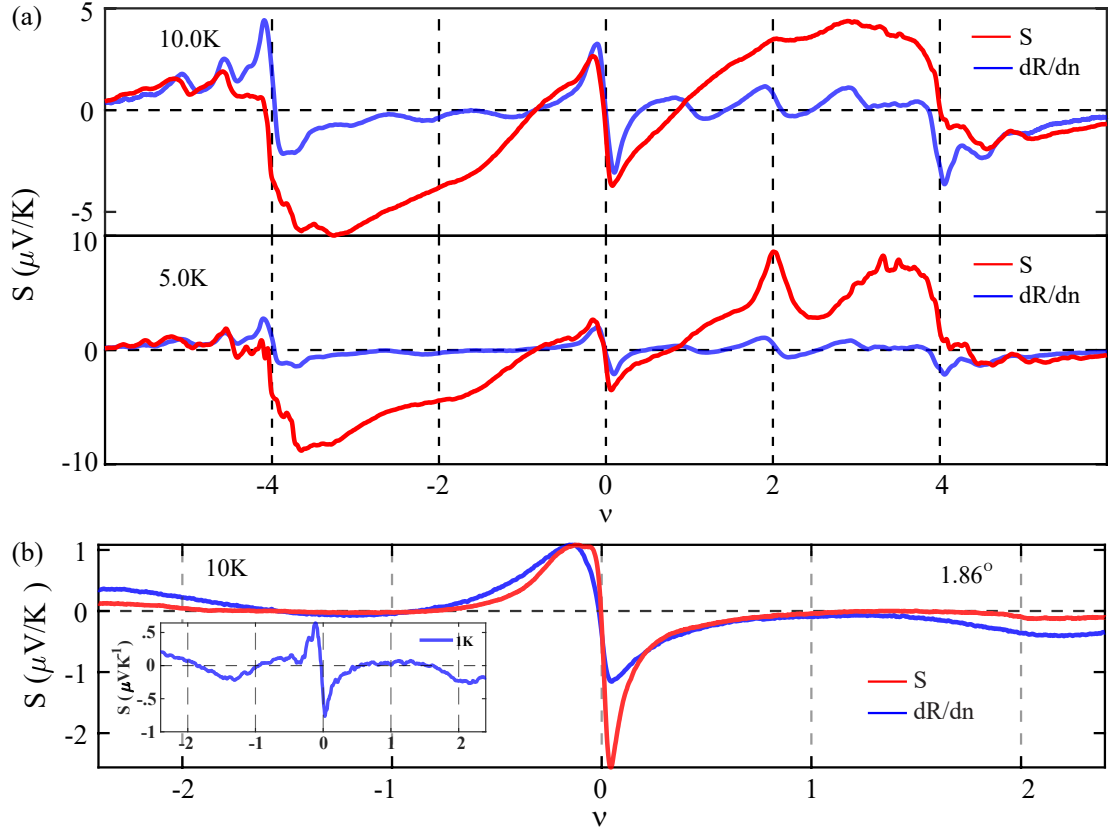

**Supplementary Figure 15:** (a) MtBLG - Comparison of measured $S$ versus $\nu$ plot (red), with $dR/dn$ versus $\nu$ plot (blue). (b) Comparison of measured $S$ with $dR/dn$ for tBLG with $1.86^o$ twist angle at $T = 10K$. The inset shows $S$ at $T = 1K$.

$\nu$ plot does not follow the sign changes of the $dR/dn$ for $\nu < \pm 4$, thus showing violation of Mott formula within the flat bands upto 10 K. However, for $\nu > \pm 4$, i.e within the dispersive bands, it can be seen that the $dR/dn$ plot even captures the fluctuation of $S$ versus $\nu$ plot (red). This shows a clear transition from interacting to non interacting regime. SI-Fig. 15(b) shows the same comparison, for the device with $1.86^o$ twist angle. However, In this case, the measured $S$ versus $\nu$ plot (red) follows the sign of $dR/dn$ versus $\nu$ plot (blue).

**SI-14 : Magnetic field dependence of thermopower for MtBLG**

SI-Fig. 16 shows the the magnetic field dependence of thermopower for the magic angle device. SI-Fig. 16(a) shows the measured $S$ versus $\nu$ as function of in-plane magnetic field for the electron side. As can be seen from the figure that while around $\nu = 1$ and 3, $S$ increases, but at $\nu = 2$ it decreases with increasing in-plane magnetic field. However, at the Dirac point ($\nu = 0$) $S$ does not change with in-plane magnetic field. In the SI-Fig. 16(c), the in-plane magnetic field dependence of $S$ for $1.86^o$ twist-angle device is shown, which remain insensitive to in-plane magnetic field. The SI-Fig. 16(b) shows the measured $S$ versus $\nu$ as a function of perpendicular magnetic field for MtBLG. The 2D plot can be seen in SI. Fig. 17c. It

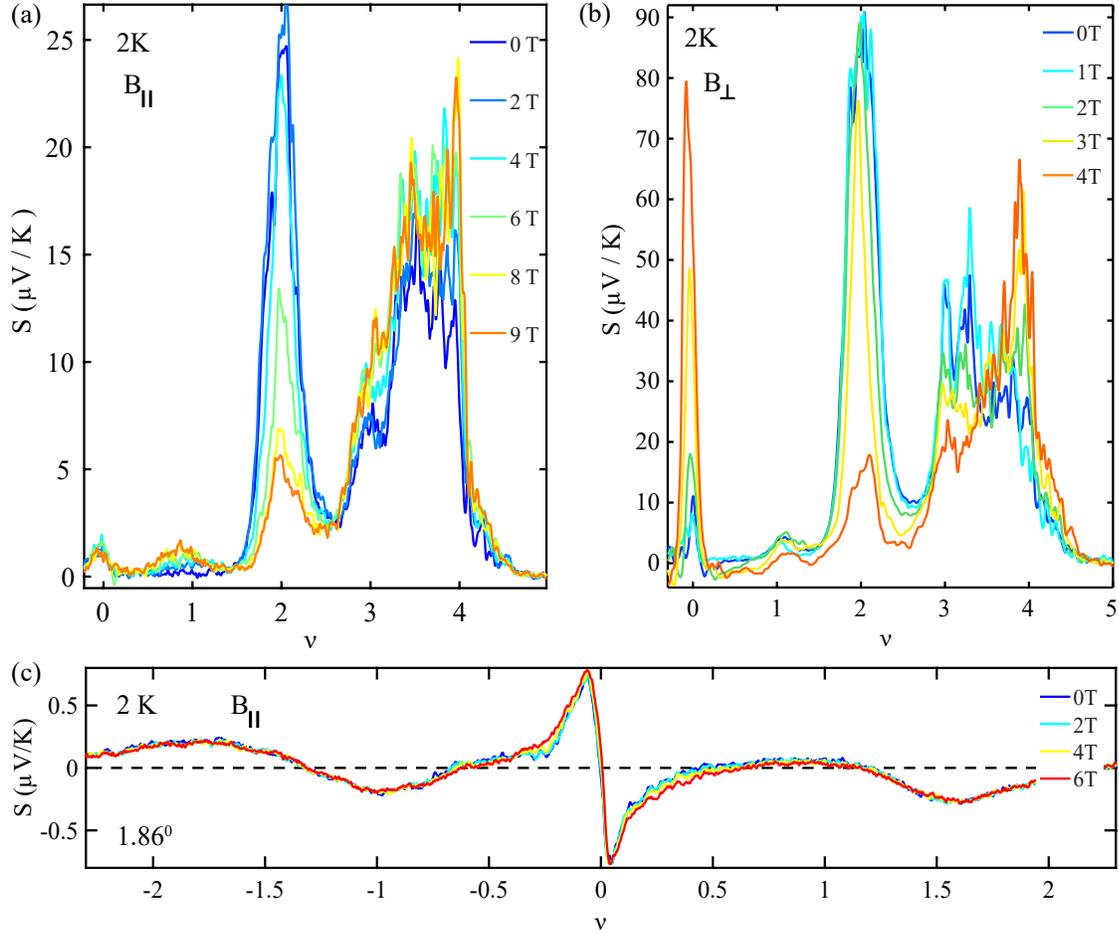

**Supplementary Figure 16:** (a) $S$ versus $\nu$ as function of in-plane magnetic field for electron side. (b) $S$ versus $\nu$ as function of perpendicular magnetic field. (c) $S$ versus $\nu$ dependence for non-magic angle device ($1.86^o$) with in-plane magnetic fields.

can seen that around $\nu = 1, 2, 3$ $S$ decreases with perpendicular magnetic field. However, around $\nu = 0$ it increases. This increase can be attributed to the appearance of Landau levels around the Dirac point, and requires further investigation.

### SI-15: Signature of Chern insulator

SI-Fig. 17a and 17b shows the response of the MtBLG device with perpendicular magnetic field. It can be seen that plateau like features arise in the resistance data above 4T. The similar plateau like features are weakly visible in the thermopower response in SI-Fig. 17c. These responses are related to the Chern insulator of MtBLG. The dashed lines correspond to Diophantine relation.

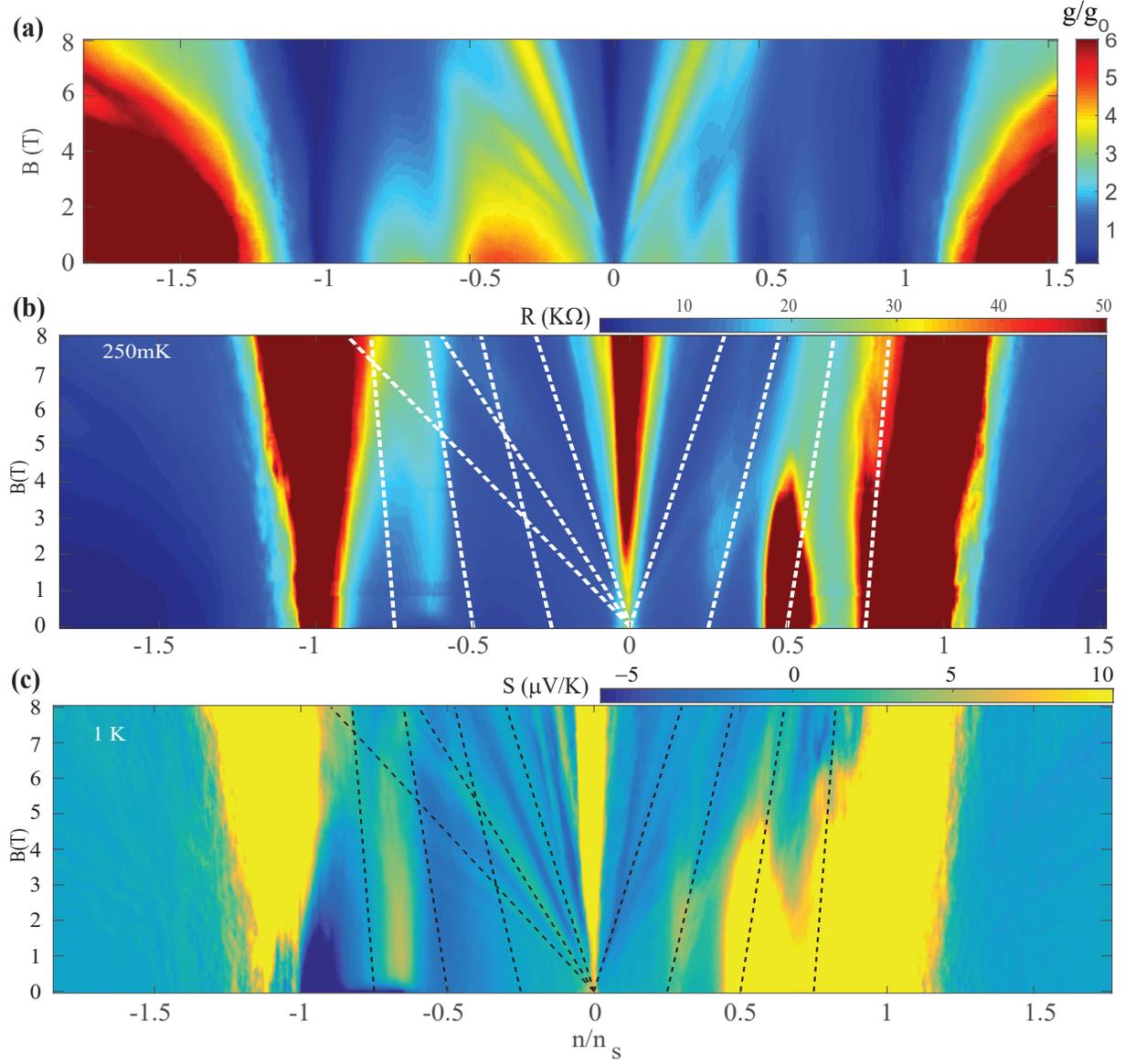

**Supplementary Figure 17:** (a) Normalized conductance with magnetic field at 2K. (b) The resistance 2D plot with filling and magnetic field. (c) 2D plot of $S$ with filling and magnetic field. The dashed lines correspond to Diophantine relation. In both the resistance and thermopower plots, the high value is saturated to visualize the plateau like features.

**SI-16: Theoretical section**

**Model:** As is Ref.[11], we use the following Hamiltonian for the MtBLG

$$\mathcal{H} = \sum_{\mathbf{k},\alpha,b} (\epsilon_{\mathbf{k}\alpha b} - \mu)\, c^\dagger_{\mathbf{k}\alpha b} c_{\mathbf{k}\alpha b} + \frac{U}{A} \sum_{\alpha < \beta} \int d^2 r\, c^\dagger_\alpha(\mathbf{r}) c^\dagger_\beta(\mathbf{r}) c_\beta(\mathbf{r}) c_\alpha(\mathbf{r}), \tag{2}$$

17

where $\alpha = 1, \ldots, 4$ are the spin-valley indices, $b$ is the moiré band index, $\mathbf{k}$ is the momentum within the first moiré Brillouin zone, $A$ is the area of the moiré unit cell, $\mu$ is the chemical potential; $c^\dagger_{\mathbf{k}\alpha b}$ and $c_{\mathbf{k}\alpha b}$ are the electron creation and annihilation operators. In the above, we have considered a local Coulomb interaction $U$ [11], with $c_\alpha(\mathbf{r}) = (1/\sqrt{N})\sum_{\mathbf{k}b} e^{i\mathbf{k}\cdot\mathbf{r}} c_{\mathbf{k}\alpha b}$ the local electron annihilation operator at $\mathbf{r}$ and $N$ is the number of moiré unit cells. We restrict the summation over the band index $b = 0, 1, 2, 3$, where 1 and 2 denote the flat bands around the moiré Dirac points, and we keep one dispersive band ($b = 0$) below and one ($b = 3$) above the flat bands. The dispersive bands are separated by a gap $\Delta = 2W$ from the flat bands (see SI-Fig.18). The latter bands are included to get a sensible behaviour of the thermopower for fillings close to the moiré flat band edges i.e. $|\nu| \sim 4$. Without these additional bands, $S(n)$ diverges at the moiré flat band edges, especially at high temperature, affecting the thermopower values even away from $|\nu| \sim 4$. Addition of the extra bands does not change the results at lower temperature $T \lesssim 0.2W \sim 10$ K. We perform Hartree-Fock approximation for the Hamiltonian (Eq.2) using the mean-field Hamiltonian $\mathcal{H}_{MF} = \sum_{\mathbf{k},\alpha,b} (\epsilon_{\mathbf{k}\alpha b} + \mu_\alpha - \mu) c^\dagger_{\mathbf{k}\alpha b} c_{\mathbf{k}\alpha b}$, where $\mu_\alpha$ are the variational parameters, and obtain the self-consistency equations

$$\mu_\alpha = U \sum_{\beta \neq \alpha} n_\beta, \tag{3a}$$

$$n_\alpha = \int_{-\infty}^{\infty} d\epsilon\, g(\epsilon) f(\epsilon + \mu_\alpha - \mu). \tag{3b}$$

Here $n_\alpha$ is the occupation of flavor $\alpha$ and $g(\epsilon) = (1/N)\sum_{\mathbf{k}b} \delta(\epsilon - \epsilon_{\mathbf{k}\alpha b})$ is the non-interacting DOS (per unit cell) which does not depend on the flavor index. $f(\epsilon) = 1/(e^{\epsilon/k_B T} + 1)$ is the Fermi function at temperature $T$. We numerically solve the above HF self-consistency equations iteratively to obtain $\mu_\alpha (n_\alpha)$ along with equation $\sum_\alpha n_\alpha = (\nu + 4) + \nu_0$ for $\mu$, where $\nu_0$ is the full filling of the dispersive band $b = 0$. We have used $U = 1.2W$ for the all the results reported below and in the main text. We have verified that our main conclusions remain valid for other values of $U \sim W$.

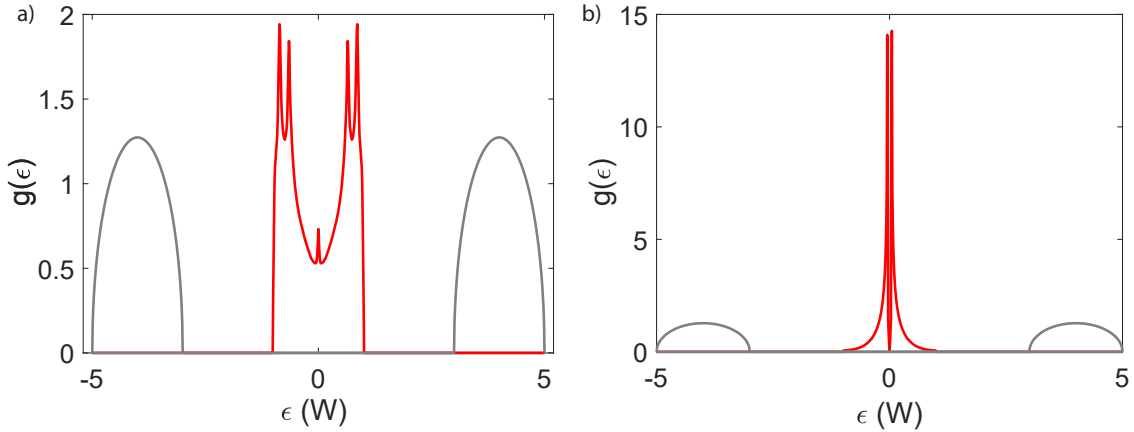

**Supplementary Figure 18: Non-interacting DOSs:** (a) DOS, $g(\epsilon)$ vs. $\epsilon$, from the continuum Bistritzer-MacDonal model [11, 12]. The semicircular DOS used in the calculations for the dispersive bands above and below the flat bands are shown. (b) DOS obtained from the continuum model with relaxation [13].

**Non-interacting DOSs:** For our HF calculations (Eq.3), we have implemented various different non-interacting DOSs $g(\epsilon)$ that have been obtained and used to describe MtBLG. For the results reported in the main text (Fig.4), we have used the DOS (referred as *continuum DOS*) obtained from the continuum Bistritzer-MacDonal model [12] in Ref.[11]. The DOS is shown in SI-Fig.18 (a) with the DOSs for the dispersive bands. Since the details of the DOS of the dispersive bands do not matter for our results, we approximate them with simple semi-circular DOS, as shown in SI-Fig. 18 (a). The continuum DOS is particle-hole (PH) symmetric with respect to the Dirac point ($\epsilon = 0$), and has two VHSs, one close to half filling, in the conduction (valence) band. The continuum DOS has a non-zero value at the Dirac point. For the continuum DOS, we find the Dirac revival transitions around the integer fillings $\nu = 1, 2, 3$ up to temperature $T \sim 0.25 - 0.3W \sim 11 - 14$ K.

We have also used the continuum model of Ref.[13] to compute the band structure and DOS (referred to as *continuum DOS with relaxation*) incorporating lattice relaxation effects for twist angle $\theta = 1.05°$. The DOS is shown in SI-Fig.18(b). The DOS is PH symmetric around $\epsilon = 0$ [$g(0) = 0$] and has sharp VHS in the conduction band at Fermi energy corresponding to $\nu \sim 1$, closer to the Dirac point, compared to the continuum DOS without relaxation. The DOS also has long tail, unlike the DOS in SI-Fig.18 (a), with $W \sim 2.5$ meV. The continuum DOS with relaxation also stabilizes the Dirac revived states with $U = 1.2W$ for all the integer fillings at least up to $T \sim 0.25W$, and specifically around $\nu = 1, 2$ up to much higher temperatures $T \sim 0.4W$. However, the signature of the Dirac revivals in transport become very weak at those high temperatures.

We have also implemented DOS obtained from tight-binding lattice model for MtBLG of Ref.[14], albeit without the lattice relaxation effects, in the HF calculations. The DOS (not shown) is PH asymmetric around $\epsilon = 0$, and has more complex arrangements of VHSs. We find the Dirac revived states to be metastable, compared to symmetry-unbroken states, for this DOS over most of the filling range. Thus we do not consider results obtained for the tight-binding DOS to compute transport coefficients.

**Transport coefficients:** We use the following expressions for resistivity $\rho$ and thermopower (Seebeck coefficient) $S$, obtained from Kubo formulae neglecting vertex corrections [15],

$$\rho = \frac{\hbar}{2\pi e^2} \frac{T}{A_0} \tag{4}$$

$$S = -\frac{k_B}{e} \frac{A_1}{A_0} \tag{5}$$

where

$$A_m = \int_{-\infty}^{\infty} d\omega \mathcal{A}^2(\epsilon, \omega) \Phi(\epsilon) \left(-T \frac{\partial f(\omega)}{\partial \omega}\right) \left(\frac{\omega}{k_B T}\right)^m, \tag{6}$$

$$\mathcal{A}(\epsilon, \omega) = -(1/\pi) \text{Im}(1/(\omega + \mu - \epsilon - i\Gamma_0)) \tag{7}$$

$$\Phi(\epsilon) = (1/NA) \sum_{\mathbf{k}\alpha b} (\partial \tilde{\epsilon}_{\mathbf{k}\alpha b}/\partial k_x)^2 \delta(\epsilon - \tilde{\epsilon}_{\mathbf{k}\alpha b}) \tag{8}$$

with $m = 1, 2$; $\mathcal{A}(\epsilon, \omega)$ is the spectral density and $\Phi(\epsilon)$ the transport DOS. Here $\Gamma_0 = 0.001W$ is small impurity broadening put in by hand to incorporate elastic scatterings, and $\tilde{\epsilon}_{\mathbf{k}\alpha b} = \epsilon_{\mathbf{k}\alpha b} + \mu_\alpha$ is the HF dispersion. For our calculations we have approximated $v^x_{\mathbf{k}\alpha b} = \partial \tilde{\epsilon}_{\mathbf{k}\alpha b}/\partial k_x$ by a constant velocity. This makes $S$ (Eq.5) independent of the band velocity and scales $\rho$ (Eq.4) by a constant factor.

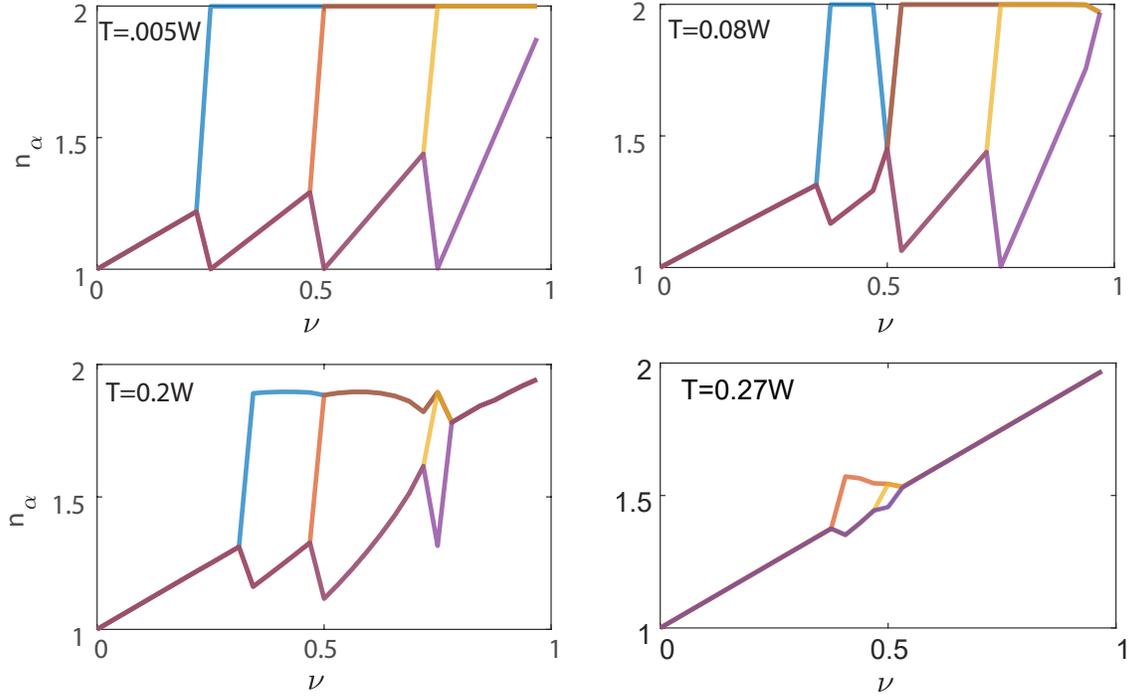

**Supplementary Figure 19: Dirac revival cascade:** The evolution of occupations $n_\alpha$ ($\alpha = 1, \ldots, 4$) of the four spin-valley flavors as a function of filling and temperature is shown for the continuum DOS [11, 12] for $U = 1.2W$.

**Thermopower from cascaded Dirac revivals:** In the Dirac revival picture for $\nu > 0$, the population ($n_\alpha$, $\alpha = 1, \ldots, 4$) of flavors follows a curious sequence of Stoner-like transitions where one, two and three of the flavors get fully occupied ($n_\alpha = 2$) at $\nu = 1, 2$ and $3$ ($n/n_s = 0.25, 0.50$ and $0.75$), respectively, resetting the fillings of the remaining flavors at $n_\alpha = 1$, as shown in SI-Fig.19 for various temperatures for the continuum DOS [11, 12]. Due to these resettings, the effective 'non-rigid' single-particle DOS undergoes successive reconstructions, as a function of both filling and temperature, leading to rather drastic PH asymmetry even at very low energies around the chemical potential ($\mu$), near each integer fillings (SI-Fig. 20). As seen from the plots of $S$ vs. $n/n_s$ in the manuscript (Fig.4c), thermopower is an extremely sensitive probe of the low-energy PH asymmetry and exhibits prominent peaks around, albeit not always exactly at, the integer fillings, over a temperature range $T \simeq 0.005W - 0.2W \sim 200$ mK $- 10$ K, for a half-bandwidth $W \simeq 4$ meV. The peak value of $S$ reaches $\sim 50 - 100$ $\mu$V/K for $n/n_s \simeq 2, 3$ $T \sim 0.1W$, consistent with our experimental observations of the manuscript (Fig.2b). For comparison, we also plot the thermopower ($S_0$) obtained using the non-interacting DOS [11, 12] in the manuscript (Fig.4c). The non-interacting thermopower, which crosses through zero at the VHS fillings, also has peaks at very low temperature ($T = 0.005W$) around the two VHSs (Fig.4c in the manuscript). But these peaks depend on the details of non-interacting DOS and are not necessarily tied to integer the fillings, unlike $S(\nu)$ in the interacting cases (Fig.4c of the manuscript). Moreover, the peaks in $S_0$ get rapidly washed away with slightly elevated temperature ($T > 0.05W$). Note that the non-interacting thermopower has been calculated

at higher temperatures ($> 100K$) using Boltzmann transport equation solver and electronic structure from an exact continuum model in Ref [16].

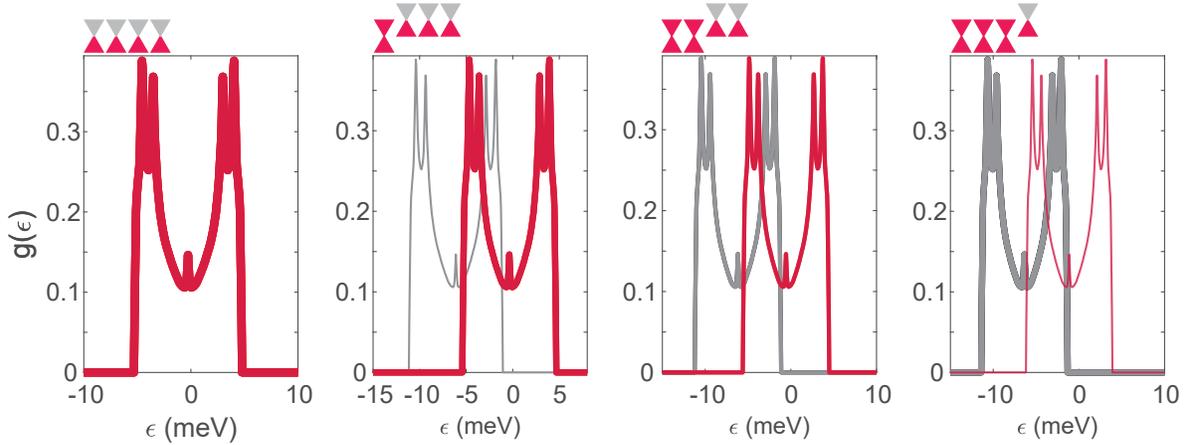

**Supplementary Figure 20:** **Evolution of effective HF DOS and emergent PH asymmetry across Dirac revivals:**. The effective DOSs obtained from HF calculation for the continuum DOS [11, 12] after each Dirac revivals around $\nu = 1, 2, 3$ are shown at low temperature $T = 0.005W$ for $U = 1.2W$. The effective DOS acquires strong low-energy PH asymmetry and leads to successive appearances of VHSs between two Dirac revivals.

**Concomitant peaks in thermopower and resistivity:** The Dirac revivals also naturally lead to peaks in resistivity (SI-Fig.21) coincident with the thermopower peaks, due to low DOS $g(\mu)$ after each Dirac resetting. The non-interacting DOS[11, 12] used for the calculations is finite at the Dirac point. As a result the Dirac revivals do not reset the DOS to zero here. However, we have verified that our main conclusions about the thermopower and resistivity peaks near integer fillings remain robust even with the continuum DOS with relaxation [13] (SI-Fig.22). In this case, the DOS vanishes at the Dirac point. Remarkably, though the system is a metal having finite DOS (Fig.4b of the manuscript) after the Dirac revivals, the resistivity $\rho(T)$ shows insulating behavior ($d\rho/dT < 0$) at low and intermediate temperature due to temperature dependence of effective DOS at a fixed filling within the HF approximation. As shown in Fig.4d of the manuscript, the temperature-dependent HF DOS also leads to non-monotonic variation of $S(T)$ with a peak around intermediate temperatures ($T \sim 0.1W$), in accordance with the experimental results in the Fig.2c of the manuscript. The spin and/or valley symmetries which are broken by the Stoner-like Dirac revival transitions are restored at higher temperature $\sim 0.3W$ over the entire filling range and the thermopower and other properties become identical to those obtained from the non-interacting DOS, as demonstrated in Fig.4c (top-panel) of the manuscript. The finite temperature transitions at a fixed filling are typically second order. However, for a range of filling near $\nu = 2$, we find that the symmetry broken state is only stabilized at finite temperature, over an intermediate range of $T$. As a result, e.g., $S(T)$ changes sign at $T \sim 0.1W$ for $\nu = 2$, as seen in the Fig.4d of the manuscript.

**Deviation of theoretically calculated thermopower from the experimental one:** Apart from the large thermopower peaks, the experimentally measured thermopower (Fig.2a of the manuscript) has positive sign

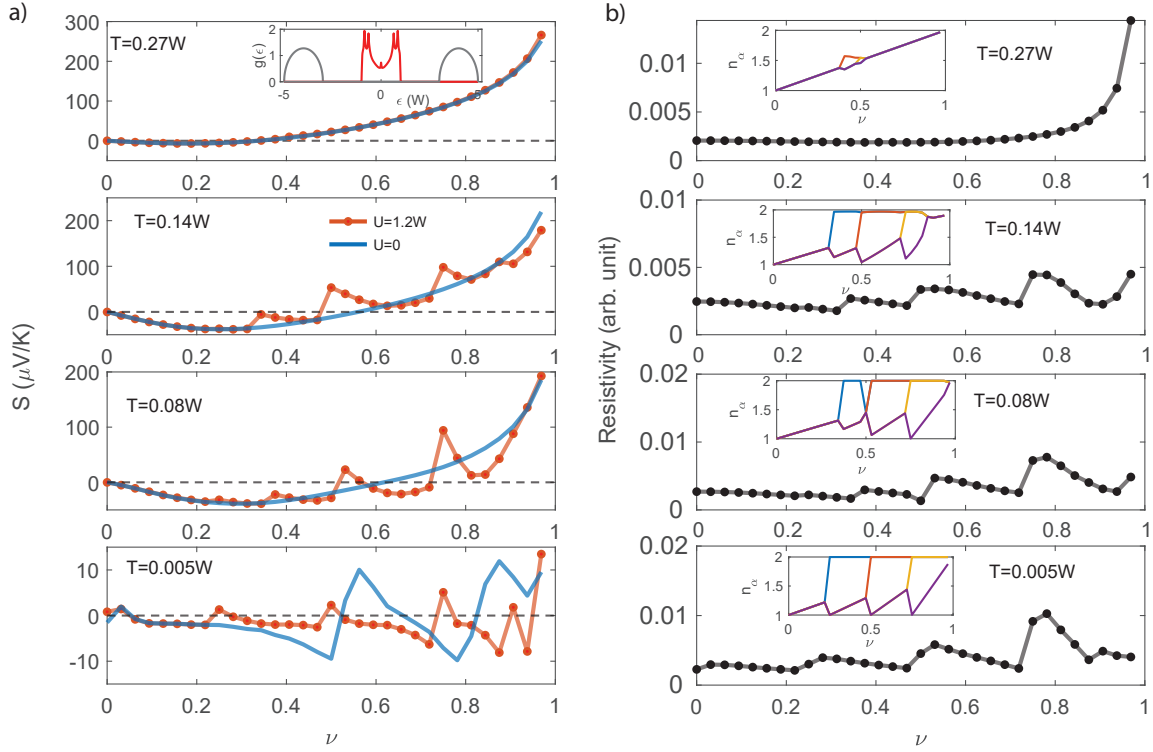

**Supplementary Figure 21: Thermopower and resistivity:** HF calculations with continuum DOS. (a) Calculated thermopower and (b) resistivity at different temperatures.

throughout the entire electron side ($\nu > 0$) for $T \lesssim 2$K. This feature is not captured by the HF calculation in the simple minimal model (Eq.2), where the effects of interaction-driven reconstruction of the DOS (Fig.20) on $S$ are only substantial close to the Dirac revivals near the integer fillings (Fig.4c of the manuscript). As a result, our theory only qualitatively captures the thermopower peaks, but the $S(n)$ follows an overall 'background' profile dictated by the non-interacting $S_0(n)$ (Fig.4c of the manuscript) and its sign change around half filling. There could be several reasons behind the deviation of theoretically calculated $S$ from the experimental one: (1) the actual single particle DOS is much more complex for MtBLG than the one we used from the simplest continuum model [11, 12, 13], (2) the twist angle inhomogeneity could be as large as 10% [11, 17] and can give broader thermopower peak around the integer fillings, thus effectively suppressing the negative contribution, and (3) away from the integer fillings the system could be a strongly correlated metal, e.g. a strange metal [18, 19]. Such correlated metallic state may arise from the Dirac revived parent states, which also lead to recurrent appearances of low-energy VHS between two successive Dirac revival transitions (SI-Fig.20). The correlation effect gets enhanced near the VHSs and lead to non-standard sign of $S$ and failure of semi-classical description for thermopower as reported recently in Ref [20]. Properties of such correlated metal cannot be captured by the HF approximation employed here.

**Effect of correlated gap in our calculation:** We also note that our theory as well as that of Ref [11] do not capture the cascaded transitions at the Dirac point ($\nu = 0$). However, large asymmetric DOS at the Dirac

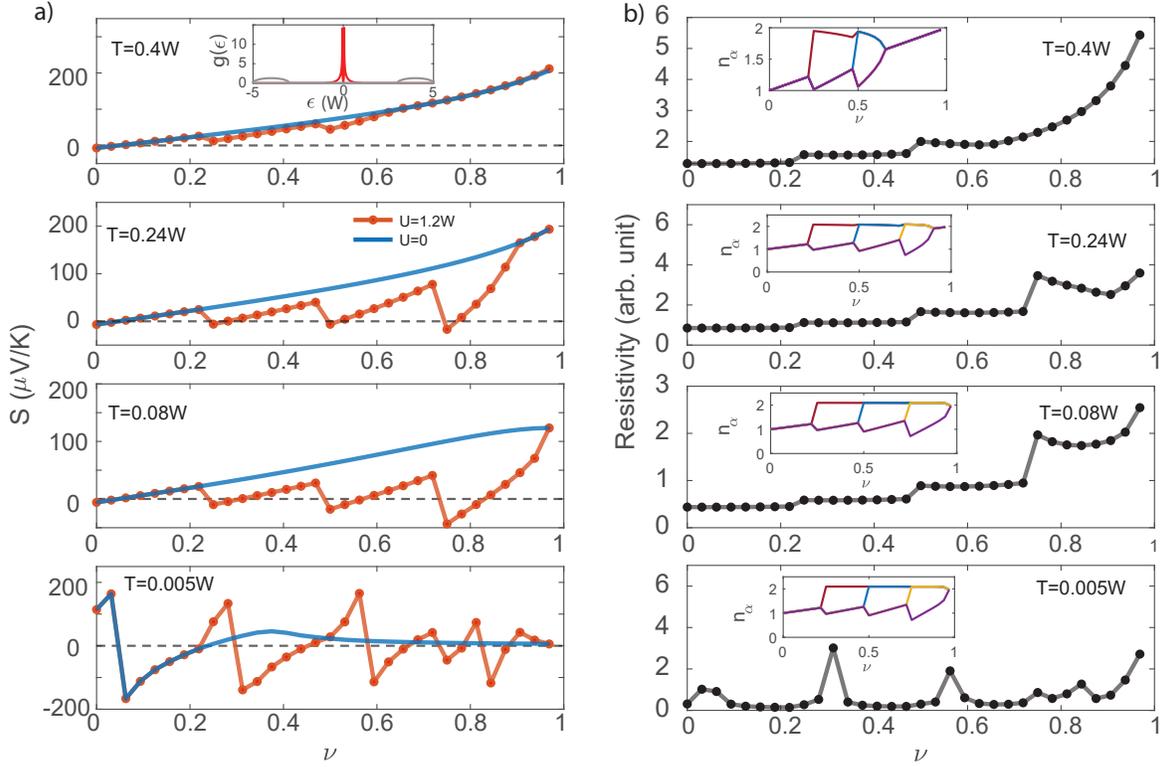

**Supplementary Figure 22: Cascade of Dirac revivals and thermopower peaks for continuum DOS with relaxation effects:** HF calculations with continuum DOS with relaxation [13], $U = 1.2W$, $W = 2.5$ meV, DOS goes to zero at the Dirac point.

point has been reported recently in STM studies [21], and thus it is not surprising to have thermopower peak at the Dirac point as seen in our experiment (Fig.4 of the manuscript). Moreover, the simplest HF approximation within the minimal model, leads to valley polarized states for the Dirac revivals at $n/n_s = 0.25, 0.75$, unlike more probable inter-valley coherent (IVC) states that have been proposed in several theoretical works [22, 23, 24]. However, such states can be easily incorporated in the model [24] and expected to lead to a small gap ($\Delta$) at the Dirac revivals. This will be consistent with the simultaneous presence of thermopower and resistance peaks at integer fillings provided $k_B T \gtrsim \Delta$. At lower temperatures, $S$ is expected to change sign across the position of resistance peak. Thus, our thermopower results put a tighter upper bound, $\Delta \sim 0.1 - 0.2$ meV (see section SI-10 and SI-11), on the correlation-induced gap at integer fillings. Below, we discuss the expected thermopower from various other kind of ground states at the integer fillings of MtBLG and its role in our measurements.

**Effect of other kind of ground states in thermopower:** The cascade of Dirac revivals (Fig. 4a) is consistent with sequence of Chern insulators seen our work and earlier studies [25] at non-zero magnetic field, which will open up gap at the Dirac point for the partially filled flavors at $\nu = 3, 2, 1$ leading to insulators with Chern numbers $C = 1, 2, 3$. The Chern insulators with large gap, however, cannot give rise to thermopower peaks at low temperature. The isospin Pomeranchuk effect, with soft spin and valley fluctuations

of the carriers, which has been observed in earlier experiments, can lead to extra entropy, and thus can enhance the thermopower at the integer fillings. However, this will be synergistic with the cascaded transitions which helps to host flavour states [26, 27], and thus enhance fluctuations. Also, we have only considered purely electronic mechanism for enhancement of thermopower. Phonon drag, where flow of phonons drags electrons via electron-phonon coupling can contribute to thermopower as well. However, this mechanism is expected to be operative only at higher temperature. For monolayer and bilayer graphene, even above $100K$ no significant contribution from phonon-drag is observed in experiment [28]. Thus, at low temperature regime $\sim 0.1K - 10K$ of our experiment the contribution from phonon-drag is expected to be negligible.

**Supplementary References**

1. Cao, Y. *et al.* Unconventional superconductivity in magic-angle graphene superlattices. *Nature* **556**, 43–50 (2018).

2. Srivastav, S. K. *et al.* Universal quantized thermal conductance in graphene. *Science Advances* **5** (2019).

3. Logvenov, G. Y., Ryazanov, V. V., Gross, R. & Kober, F. Comment on "anomalous peak in the thermopower of $yba_2cu_3o_{7-\delta}$ single crystals: A possible fluctuation effect". *Phys. Rev. B* **47**, 15322–15323 (1993).

4. Fong, K. C. & Schwab, K. Ultrasensitive and wide-bandwidth thermal measurements of graphene at low temperatures. *Physical Review X* **2**, 031006 (2012).

5. Crossno, J. *et al.* Observation of the dirac fluid and the breakdown of the wiedemann-franz law in graphene. *Science* **351**, 1058–1061 (2016).

6. Betz, A. *et al.* Supercollision cooling in undoped graphene. *Nature Physics* **9**, 109–112 (2013).

7. Talanov, A. V., Waissman, J., Taniguchi, T., Watanabe, K. & Kim, P. High-bandwidth, variable-resistance differential noise thermometry. *Review of Scientific Instruments* **92**, 014904 (2021).

8. Aamir, M. A. *et al.* Ultrasensitive Calorimetric Measurements of the Electronic Heat Capacity of Graphene. *Nano Letters* **21**, 5330–5337 (2021).

9. Polshyn, H. *et al.* Large linear-in-temperature resistivity in twisted bilayer graphene. *Nature Physics* **15**, 1011–1016 (2019).

10. Cao, Y. *et al.* Nematicity and competing orders in superconducting magic-angle graphene. *Science* **372**, 264–271 (2021).

11. Zondiner, U. *et al.* Cascade of phase transitions and dirac revivals in magic-angle graphene. *Nature* **582**, 203–208 (2020).

12. Bistritzer, R. & MacDonald, A. H. Moiré bands in twisted double-layer graphene. *Proceedings of the National Academy of Sciences* **108**, 12233–12237 (2011).

13. Koshino, M. *et al.* Maximally localized wannier orbitals and the extended hubbard model for twisted bilayer graphene. *Phys. Rev. X* **8**, 031087 (2018).

14. Moon, P. & Koshino, M. Energy spectrum and quantum hall effect in twisted bilayer graphene. *Phys. Rev. B* **85**, 195458 (2012).

15. Pálsson, G. & Kotliar, G. Thermoelectric response near the density driven mott transition. *Phys. Rev. Lett.* **80**, 4775–4778 (1998).

16. Kommini, A. & Aksamija, Z. Very high thermoelectric power factor near magic angle in twisted bilayer graphene. *2D Materials* **8**, 045022 (2021).

17. Uri, A. *et al.* Mapping the twist-angle disorder and Landau levels in magic-angle graphene. *Nature* **581**, 47–52 (2020).

18. Cao, Y. *et al.* Strange metal in magic-angle graphene with near planckian dissipation. *Phys. Rev. Lett.* **124**, 076801 (2020).

19. Jaoui, A. *et al.* Quantum-critical continuum in magic-angle twisted bilayer graphene. *arXiv:2108.07753* (2021).

20. Ghawri, B. *et al.* Excess entropy and breakdown of semiclassical description of thermoelectricity in twisted bilayer graphene close to half filling. *arXiv preprint arXiv:2004.12356* (2020).

21. Choi, Y. *et al.* Correlation-driven topological phases in magic-angle twisted bilayer graphene. *Nature* **589**, 536–541 (2021).

22. Po, H. C., Zou, L., Vishwanath, A. & Senthil, T. Origin of mott insulating behavior and superconductivity in twisted bilayer graphene. *Phys. Rev. X* **8**, 031089 (2018).

23. Bultinck, N. *et al.* Ground state and hidden symmetry of magic-angle graphene at even integer filling. *Phys. Rev. X* **10**, 031034 (2020).

24. Shavit, G., Berg, E., Stern, A. & Oreg, Y. Theory of correlated insulators and superconductivity in twisted bilayer graphene. *arXiv:2107.08486* (2021).

25. Andrei, E. Y. & MacDonald, A. H. Graphene bilayers with a twist. *Nature Materials* **19**, 1265–1275 (2020).

26. Rozen, A. *et al.* Entropic evidence for a Pomeranchuk effect in magic-angle graphene. *Nature* **592**, 214–219 (2021).

27. Saito, Y. *et al.* Isospin pomeranchuk effect in twisted bilayer graphene. *Nature* **592**, 220–224 (2021).

28. Zuev, Y. M., Chang, W. & Kim, P. Thermoelectric and magnetothermoelectric transport measurements of graphene. *Phys. Rev. Lett.* **102**, 096807 (2009).


\end{document}